\documentclass[paper]{JHEP3}
\pdfoutput=1
\usepackage{amsmath,amssymb,amsthm,amscd,graphicx}

\input epsf.sty

\theoremstyle{definition}

\newcommand{\CF}{{\cal F}}

\newcommand{\CH}{{\cal H}}

\newcommand{\CN}{{\cal N}}
\newcommand{\CO}{{\cal O}}

\newcommand{\CQ}{{\cal Q}}

\def\IS{{\mathbb S}}

\newcommand{\re}{{\rm e}}
\newcommand{\ri}{{\rm i}}
\newcommand{\rd}{{\rm d}}

\newcommand{\Ai}{\mathop{\rm Ai}\nolimits}

\newcommand{\wigner}{{\mathrm{W}}}

\newcommand{\starprod}{\mathop{\overrightarrow{{\bigstar}}}}

\newcommand{\be}{\begin{equation}}
\newcommand{\ee}{\end{equation}}
\newcommand{\ba}{\begin{aligned}}
\newcommand{\ea}{\end{aligned}}
\newcommand{\ben}{\begin{eqnarray}\displaystyle}
\newcommand{\een}{\end{eqnarray}}

\newcommand{\sectiono}[1]{\section{#1}\setcounter{equation}{0}}

\newdimen\tableauside\tableauside=1.0ex
\newdimen\tableaurule\tableaurule=0.4pt
\newdimen\tableaustep
\def\phantomhrule#1{\hbox{\vbox to0pt{\hrule height\tableaurule width#1\vss}}}
\def\phantomvrule#1{\vbox{\hbox to0pt{\vrule width\tableaurule height#1\hss}}}
\def\sqr{\vbox{%
  \phantomhrule\tableaustep
  \hbox{\phantomvrule\tableaustep\kern\tableaustep\phantomvrule\tableaustep}%
  \hbox{\vbox{\phantomhrule\tableauside}\kern-\tableaurule}}}
\def\squares#1{\hbox{\count0=#1\noindent\loop\sqr
  \advance\count0 by-1 \ifnum\count0>0\repeat}}
\def\tableau#1{\vcenter{\offinterlineskip
  \tableaustep=\tableauside\advance\tableaustep by-\tableaurule
  \kern\normallineskip\hbox
    {\kern\normallineskip\vbox
      {\gettableau#1 0 }%
     \kern\normallineskip\kern\tableaurule}%
  \kern\normallineskip\kern\tableaurule}}
\def\gettableau#1{\ifnum#1=0\let\next=\null\else
\squares{#1}\let\next=\gettableau\fi\next}

\newcommand{\figref}[1]{Fig.~\protect\ref{#1}}

\title{Interacting fermions and ${\cal N}=2$ Chern--Simons--matter theories}

\author{
Marcos Mari\~no and Pavel Putrov
\\
D\'epartement de Physique Th\'eorique et Section de Math\'ematiques,\\
Universit\'e de Gen\`eve, Gen\`eve, CH-1211 Switzerland\\
\\
\email{marcos.marino@unige.ch}, \quad
\email{pavel.putrov@unige.ch}
}

\abstract{The partition function on the three-sphere of $\CN=3$ Chern--Simons--matter theories can be formulated in terms of an ideal Fermi gas. In this paper we show that, in theories with $\CN=2$ supersymmetry, the partition function corresponds to a gas of interacting fermions in one dimension. The large $N$ limit is the thermodynamic limit of the gas and it can be analyzed with the Hartree and Thomas--Fermi approximations, which lead to the known large $N$ solutions of these models. We use this interacting fermion picture to analyze in detail $\CN=2$ theories with one single node. In the case of theories with no long-range forces we incorporate exchange effects and argue that the partition function is given by an Airy function, as in $\CN=3$ theories. For the theory with $g$ adjoint superfields and long-range forces, the Thomas--Fermi approximation leads to an integral equation which determines the large $N$, strongly coupled $R$-charge.  
}

\begin{document}

\sectiono{Introduction}

In the last years, the use of localization techniques in the study of Chern--Simons--matter (CSM) theories has led to a new class of matrix models which generalize the matrix models for pure Chern--Simons theory introduced in \cite{lr,mmcs}. These models compute partition functions and Wilson loop correlators on the three-sphere and other compact three-manifolds. They were first introduced in \cite{kwy} for CSM theories with ${\cal N}\ge 3$ supersymmetry, and the result was extended to theories with ${\cal N}=2$ supersymmetry in \cite{jafferis,hama}, see \cite{lectures} for a review and a list of references. 
The study of these matrix models has led to many interesting results. It has provided precision tests of the AdS$_4$/CFT$_3$ correspondence as well as beautiful 
field-theoretical results on superconformal field theories in three dimensions. It is therefore interesting to find efficient ways to analyze these models, in particular in the large $N$ limit, where 
they make contact with superstrings and M-theory. 

The matrix model corresponding to ABJM theory was solved in the 't Hooft expansion, at all orders in $1/N$, in \cite{mpwilson,dmp}, by using techniques developed earlier in random matrix theory and topological string theory. However, these standard techniques seem to be of limited use for models with $\CN \le 3$ supersymmetry (see however \cite{cmp,suyama} for planar solutions of some of these models by using those techniques). For more general theories with $\CN=3$ supersymmetry, \cite{hklebanov} developed a powerful technique to study the so-called M-theory limit, where $N$ is large and the coupling constants are fixed. This technique can be generalized to $\CN=2$ theories and it has made possible to understand the leading, large $N$ limit of the free energy of these matrix models, see \cite{jkps,ms,k1,ghp1,ghp2} for a non-exhaustive list of examples.

In \cite{mp}, a different framework was introduced to analyze $\CN=3$ theories, based on a reformulation of the matrix model as the partition function of an ideal Fermi gas. This reformulation has various virtues: first of all, it gives an elementary derivation and a nice physical picture of the $N^{3/2}$ behavior of the free energy of these models in terms of free fermions. Second, it makes possible to compute all $1/N$ corrections to the free energy. This leads to a simple derivation of the Airy function result for the partition function of ABJM theory \cite{fhm}. In fact, it was conjectured in \cite{mp} that in theories displaying the $N^{3/2}$ behavior of the free energy, the partition function at all orders in $1/N$ is given by an Airy function, and the Fermi gas reformulation makes possible to establish this conjecture for many $\CN=3$ theories. The Fermi gas picture also leads to powerful results for the Wilson loop vevs of ABJM theory at all orders in $1/N$ \cite{kmss}, and it is natural to see how it can be generalized and/or used in other situations. 

In this paper we generalize some of the results of \cite{mp} to CSM theories with $\CN=2$ supersymmetry. 
We find that the partition function on the three-sphere of these theories can be reformulated, via the matrix model of 
\cite{kwy,jafferis,hama}, as the partition function of an {\it interacting} Fermi gas. For theories with $\CN\ge 3$ supersymmetry, the Fermi gas is non-interacting but there is a non-trivial external potential, and all the physics reduces to one-body physics. In the case of $\CN=2$ supersymmetry, we have in general two-body and higher body interactions. Interacting quantum gases are notoriously hard to analyze, and this one is not an exception. However, in the thermodynamic limit (which corresponds to the large $N$ limit of the matrix model), one can use the 
Hartree/Thomas-Fermi approximation. We show that this approximation leads to the large $N$ treatment of the $\CN=2$ theories proposed in 
\cite{jkps,ms,k1}. Unfortunately, a systematic understanding of the $1/N$ corrections to this leading, large $N$ 
result seems difficult in general. In the cases of flavored theories with one node first considered in 
\cite{benini,jaff-flav}, it is in principle possible to 
go beyond the Thomas--Fermi approximation, and we compute the next-to-leading correction due to exchange effects in the interacting Fermi gas. We also study a theory with one node but with long-range interactions between the fermions, namely the CS theory with adjoint multiplets considered in \cite{gy,minwalla, niarchos, mn}. In this case, the long-range forces lead to a qualitatively different physics for the interacting fermions, which can be still studied in detail at large $N$ by using the Thomas--Fermi approximation. We obtain in particular an 
integral equation which determines the large $N$, strongly coupled $R$-charge of the multiplets.   

The organization of this paper is the following. In section 2 we review the construction of matrix models for $\CN=2$ CSM theories, we show how to interpret them in terms of interacting fermions, and 
we explain how to obtain the large $N$ solution via the Hartree/Thomas--Fermi approximation, reproducing in this way the density functional approach of \cite{hklebanov,jkps} and other papers. In section 3 we study in detail one node, flavored theories. We analyze them in terms of an interacting Fermi gas, we solved them with the Thomas--Fermi approximation, recovering the solution of \cite{jkps}, and we calculate the next-to-leading, exchange correction to this leading result. We argue that the partition function is an Airy function. In section 4, we study from the point of view of the interacting Fermi gas a one-node theory with long-range forces, namely CS theory with adjoint multiplets, and we derive an integral equation for the large $N$ limit of the R-chage. This allows us to re-derive, and partially improve,  some results in \cite{minwalla} on this theory. Finally, we conclude with some remarks and open problems.

\sectiono{$\CN=2$ Chern--Simons--matter theories as interacting Fermi gases}

\subsection{Matrix models for $\CN=2$ Chern--Simons--matter theories}
\label{matrix-rules}
In \cite{kwy} it was shown, by using localization techniques first introduced in this context in \cite{pestun}, that the partition function of $\CN\ge 3$ Chern--Simons--matter 
theories can be written as a matrix integral. This result was generalized to $\CN=2$ theories in \cite{jafferis, hama}. We now give a brief summary of the 
ingredients involved in these $\CN=2$ matrix models. A basic building block is the function 
 \begin{equation} \ell(z) = - z
\log\left(1-\re^{2\pi \ri z}\right) + \frac{\ri}{2} \left(\pi z^2 +
\frac{1}{\pi} \textrm{Li}_2(\re^{2\pi \ri z}) \right) - \frac{\ri
\pi}{12}, 
\end{equation}
which has the following basic properties: 
\begin{enumerate}
\item It is odd
\be
\ell(z)=-\ell(-z).
\ee
\item Using the standard expansions of the logarithm and dilogarithm, one immediately shows that $\ell(z)$ has the asymptotic expansion 
\be
\label{asyml}
\ell(z) = \pm {\ri \pi \over 2} \left( z^2-{1\over 6} \right) +\sum_{m=1}^{\infty} \left( {z\over m} \pm {\ri \over 2 \pi m^2} \right) \re^{\pm 2 \pi \ri m z},
\ee
where the $\pm$ sign corresponds to ${\rm Im}(z)\gg 1$ or ${\rm Im}(z)\ll -1$, respectively. 

\item It satisfies the equation 
\be
{\rd \ell \over \rd z} = -\pi z \cot(\pi z). 
\ee
\item If we denote 
\be
z=\widetilde \Delta + {\ri \theta \over 2 \pi}
\ee
and
\be
\widetilde \Delta= 1-\Delta
\ee
we have from (\ref{asyml}) 
\be
\label{largeDelta}
\ell(z)+ \ell (z^*)=-\widetilde \Delta |\theta| - v_\Delta (\theta),
\ee
where the function $v_\Delta (\theta)$ has the following expansion 
\be
\label{srp}
v_\Delta (\theta)=
\sum_{m=1}^\infty \re^{-m|\theta|}  \left[ \left( {|\theta| \over 
 \pi m} +{1\over \pi m^2} \right)\sin\left(2 \pi m \widetilde \Delta\right) -{2 \widetilde\Delta \over m} \cos\left( 2 \pi m \widetilde \Delta\right) \right], \qquad |\theta| \gg 1. 
\ee
In this paper, this function will be interpreted as a short-range potential. 

\item With the above notations, if $\widetilde \Delta=1/2$, we have
\be
\ell(z) + \ell(z^*)=-\log \left[ 2 \cosh{\theta \over 2} \right]. 
\ee

\item It is not difficult to show that the Fourier transform of $v_\Delta(x)$ is given by the simple expression, 
\be
\label{ft}
\widehat v_\Delta(\omega)= {{\sqrt{2 \pi}} \over \omega }  \left[ {1-\Delta \over  \pi \omega} + {\sinh\left( 2 \pi (\Delta-1)  \omega  \right) \over \cosh(2 \pi  \omega)-1} \right], 
\ee
where the Fourier transform of a function $f(x)$ is defined as
\be
\widehat f(\omega)={1\over {\sqrt{2 \pi}}} \int_{-\infty}^\infty f(x) \re^{\ri \omega x} \rd x.
\ee

\item Let $f(x)$ be a function. Up to exponentially suppressed terms, we have
\be
\label{approxker}
\int_{-\infty}^\infty \rd x' \, v_\Delta(x-x') f(x') \approx \left( \int_{-\infty}^\infty \rd x' \, v_\Delta(x') \right) f(x).
\ee
The integral appearing above can be computed by using the Fourier transform (\ref{ft}): 
\be
\label{vint}
 \int_{-\infty}^\infty \rd x \, v_\Delta(x)  ={\sqrt{2 \pi}} \widehat v_\Delta(0)= {2  \pi^2 \over 3} (\Delta-1) \left( 2 \Delta^2-4 \Delta+1\right). 
\ee
\end{enumerate}

We now consider a general $\CN=2$ quiver CSM theory, made up of nodes connected by edges. Each node 
has a $U(N)$ Chern--Simons gauge theory with level $k_a$, where $a=1, \cdots, r$ labels the nodes. 
Each node has an associated to it set of $N$ eigenvalues $\lambda_i^{(a)}$, $i=1, \cdots, N$, and the matrix integral is obtained by integrating 
over all the eigenvalues. We will now list the contribution of the different fields to the integrand of the matrix integral. We denote by $\lambda_i$ the variables corresponding to the $a$ node,
and by $\mu_i$ those corresponding to the $b$ node. 

The contribution of the CS vector multiplet at the node $a$ gives a factor
\be
\prod_{i<j} \left( 2 \sinh  {\lambda_i-\lambda_j  \over 2} \right)^2, 
\ee
while the classical CS action leads to 
\be
\exp\left( {\ri k_a \over 4 \pi} \sum_{i=1}^N \left(\lambda_i\right)^2\right). 
\ee
We will assume, as in \cite{jkps}, that the nodes are connected by pairs of bifundamental fields $A_{ab}$, $B_{ba}$, with anomalous dimensions $\Delta_{(a,b)}$ and $\Delta_{(b,a)}$, respectively. 
This leads to the following factor in the integrand 
\be
\prod_{ij} \exp \left[ \ell\left( \widetilde \Delta_{(a,b)} + \ri {\lambda_i -\mu_j \over 2 \pi} \right) + \ell \left( \widetilde \Delta_{(b,a)} - \ri {\lambda_i -\mu_j \over 2 \pi} \right)\right]. 
\ee
A field in the adjoint representation in the $a$-th node is represented by 
\be
\label{adj}
\prod_{ij} \exp \left[ \ell\left( \widetilde \Delta_{a} + \ri {\lambda_i -\lambda_j \over 2 \pi} \right)\right],
\ee
while a field in the (anti) fundamental gives
\be
\prod_{ij} \exp \left[ \ell\left( \widetilde \Delta_{f_a} \pm  \ri {\lambda_i \over 2 \pi} \right)\right].
\ee

\subsection{The interacting Fermi gas picture}

In \cite{mp} it was shown that, in the case of $\CN=3$ Chern--Simons--matter theories, the matrix integral obtained from the localization approach of \cite{kwy,jafferis,hama} can be re-expressed as the partition function of a quantum, one-dimensional, non interacting Fermi gas. The number of fermions, $N$, is simply the rank of the gauge group $U(N)$. Let us review this result by using a formalism suitable for generalizations. We will consider the generalization of ABJM theory given by necklace quivers with $r$ nodes\cite{quiver1,quiver2}, and with fundamental matter in each node. These theories have a gauge group
\be
U(N)_{k_1} \times U(N)_{k_2} \times \cdots U(N)_{k_r}
\ee
and each node will be labelled with the letter $a=1, \cdots, r$. There are bifundamental chiral superfields $A_{a a+1}$, $B_{a a-1}$ connecting
adjacent nodes, and in addition we will suppose that there are $N_{f_a}$ matter superfields $(Q_a, \tilde Q_a)$ in each node, in the fundamental representation. We will write
\be
\label{CSlevels}
k_a=n_a k,
\ee
and we will assume that
\be
\label{add0}
\sum_{a=1}^r n_a=0.
\ee

According to the rules reviewed above, the matrix model computing the $\IS^3$ partition function of this necklace quiver is given by
\be
\label{quivermm}
Z(N)={1\over (N!)^r} \int  \prod_{a,i} {\rd \lambda_{a,i} \over 2 \pi}  {\exp \left[ {\ri n_a k\over 4 \pi}\lambda_{a,i}^2 \right] \over \left( 2 \cosh{\lambda_{a,i} \over 2}\right)^{N_{f_a}} } \prod_{a=1}^r  {\prod_{i<j} \left[ 2 \sinh \left( {\lambda_{a,i} -\lambda_{a,j} \over 2} \right)\right]^2 \over \prod_{i,j} 2 \cosh \left( {\lambda_{a,i} -\lambda_{a+1,j} \over 2} \right)}.
\ee
The building block of the integrand in (\ref{quivermm}) is the following $N$-dimensional kernel, associated to an edge
connecting the nodes $a$ and $b$:
\be
\label{multiK}
K_{ab}(\lambda_1, \cdots, \lambda_N; \mu_1, \cdots, \mu_N)= {1\over N!} \prod_{i=1}^N \re^{-U_a(\lambda_i)} {\prod_{i<j}  2 \sinh \left( {\lambda_{i} -\lambda_{j} \over 2 } \right)  2 \sinh \left( {\mu_{i} -\mu_{j} \over 2 } \right)\over \prod_{i,j} 2 \cosh \left( {\lambda_{i} -\mu_j \over 2 } \right)}.
\ee
Here,
\be
\label{one-body}
U_a(\lambda)=-{\ri n_a k \over 4 \pi }\lambda^2 +N_{f_a} \log\left( 2 \cosh{\lambda \over 2 }\right)
\ee
and it will be interpreted as a one-body potential for a Fermi gas with $N$ particles. 

To make the connection with a Fermi gas, we introduce the projection operator on totally antisymmetric states
\be
P={1\over N!} \sum_{\sigma \in S_N} (-1)^{\epsilon(\sigma)} \sigma, 
\ee
which satisfies 
\be
P^2=P. 
\ee
Let 
\be
| \lambda_1, \cdots, \lambda_N \rangle
\ee
be the basis of space eigenstates for an $N$-particle system $\CH_N$. We introduce the appropriately antisymmetrized states
\be
\left| \lambda_1, \cdots, \lambda_N \right \} =P | \lambda_1, \cdots, \lambda_N \rangle={1\over N!}  \sum_{\sigma \in S_N} (-1)^{\epsilon(\sigma)} 
| \lambda_{\sigma(1)}, \cdots, \lambda_{\sigma(N)} \rangle
\ee
which are a basis of the Hilbert space of fermions $\CF_N$ (see chapter 1 of \cite{no} for a very useful summary of these properties). We now want to interpret the 
kernel  (\ref{multiK}) as a matrix element
\be
K_{ab}(\lambda_1, \cdots, \lambda_N; \mu_1, \cdots, \mu_N)= \left\{ \lambda_1, \cdots, \lambda_N \right| \hat \rho_{ab} \left| \mu_1, \cdots, \mu_N\right\},
\ee
in terms of a non-symmetrized density matrix $\hat \rho_{ab}$ (i.e. a density matrix for distinguishable particles). We first notice that
\be
\label{symrhoK}
  \left\{ \lambda_1, \cdots, \lambda_N \right| \hat \rho_{ab} \left| \mu_1, \cdots, \mu_N\right\}={1\over N!} \sum_{\sigma \in S_N} (-1)^{\epsilon(\sigma)}
\rho_{ab} \left( \lambda_1, \cdots, \lambda_N; \mu_{\sigma(1)}, \cdots, \mu_{\sigma(N)}\right).
\ee
We now use the Cauchy identity
 \be
 \label{cauchy}
 \ba
  {\prod_{i<j}  \left[ 2 \sinh \left( {\mu_i -\mu_j \over 2}  \right)\right]
\left[ 2 \sinh \left( {\nu_i -\nu_j   \over 2} \right) \right] \over \prod_{i,j} 2 \cosh \left( {\mu_i -\nu_j \over 2} \right)}
 & ={\rm det}_{ij} \, {1\over 2 \cosh\left( {\mu_i - \nu_j \over 2} \right)}\\
 &=\sum_{\sigma \in S_N} (-1)^{\epsilon(\sigma)} \prod_i {1\over 2 \cosh\left( {\mu_i - \nu_{\sigma(i)} \over 2} \right)}.
 \ea
  \ee
  In this equation, $S_N$ is the permutation group of $N$ elements, and $\epsilon(\sigma)$ is the signature of the permutation $\sigma$.
  We obtain,
\be
\ba
K_{ab}(\lambda_1, \cdots, \lambda_N; \mu_1, \cdots, \mu_N)&={1\over N!} \prod_{i=1}^N \re^{-U_a(\lambda_i)} {\rm det}_{ij} \left( {1\over 2 \cosh {\lambda_i-\mu_j\over 2}} \right)\\
&=
{1\over N!} \sum_{\sigma \in S_N} (-1)^{\epsilon(\sigma)} \prod_{i=1}^N \re^{-U_a(\lambda_i)} \prod_{i=1}^N t(\lambda_i-\mu_{\sigma(j)}),
\ea
\ee
where we denoted
\be
\label{tx}
t(x)={1\over 2 \cosh {x\over 2}}.
\ee
By comparing with (\ref{symrhoK}), it follows that
\be
\rho_{ab}\left( \lambda_1, \cdots, \lambda_N; \mu_1, \cdots, \mu_N\right)=\prod_{i=1}^N \re^{-U_a(\lambda_i)} \prod_{i=1}^N t(\lambda_i- \mu_i).
\ee
This factorization tells us that the $N$-particle system is non-interacting, since the density matrix is completely factorized. 

A more general construction is possible which takes into account further interactions between the nodes. In fact, {\it any} kernel $K$ which is antisymmetric in $\lambda_i$, $\mu_j$ defines a density matrix $\hat \rho$ through the equation 
\be
K\left( \lambda_1, \cdots, \lambda_N; \mu_1, \cdots, \mu_N\right)= \left\{ \lambda_1, \cdots, \lambda_N \right| \hat \rho \left| \mu_1, \cdots, \mu_N\right\}. 
\label{K-rho-relation}
\ee
Let us then assume that we have a more general kernel, of the form 
\be
\ba
\label{genK}
&K_{ab}(\lambda_1, \cdots, \lambda_N; \mu_1, \cdots, \mu_N)\\
&={1\over N!} \prod_{i=1}^N \re^{-U_a(\lambda_i)} \prod_{i,j} \re^{-f_{ab} (\lambda_i, \mu_j)} \prod_{i,j} \re^{-W_b(\mu_i, \mu_j)}{\rm det}_{ij} \left( {1\over 2 \cosh {\lambda_i-\mu_j\over 2}} \right), 
\ea
\ee
which includes a two-body interaction in the node $b$, given by the potential $W_b$, as well as a general interaction between the two nodes $a$, $b$, 
$f_{ab}$. We first notice that 
\be
\prod_{i,j} \re^{-f_{ab} (\lambda_i, \mu_j)} \prod_{i,j} \re^{-W_b(\mu_i, \mu_j)}
\ee
are invariant under any permutation of the $\lambda_i$, $\mu_i$ (since we are taking a product over all possible pairs of particles). 
Therefore, the r.h.s. of (\ref{genK}) defines an $N$-dimensional kernel $ \rho_{ab}$. We can write 
\be
\rho_{ab}(\lambda_1, \cdots, \lambda_N; \mu_1, \cdots, \mu_N)=\prod_{i=1}^N \re^{-U_a(\lambda_i)} \prod_{i,j} \re^{-f_{ab} (\lambda_i, \mu_j)} \prod_{i,j} \re^{-W_b(\mu_i, \mu_j)}\prod_i t(\lambda_i- \mu_i). 
\ee
We will assume, as it happens in the $\CN=2$ matrix integrals, that 
\be
f_{ab}(\lambda,\mu)=f_{ab}(\lambda-\mu), \qquad  W_b(\mu,\mu')=W_b(\mu-\mu').
\ee
We can then write, 
\be
\rho_{ab}(\lambda_1, \cdots, \lambda_N; \mu_1, \cdots, \mu_N)=\prod_{i=1}^N \re^{-U_a(\lambda_i)} g_{ab} (\lambda_i-\mu_i) \prod_{i\not=j} \re^{-f_{ab} (\lambda_i-\mu_j)} \prod_{i,j} \re^{-W_b(\mu_i-\mu_j)},
\ee
where 
\be
\label{gfx}
g_{ab}(y)=\re^{ -f_{ab}(y)} t(y)
\ee
incorporates the diagonal elements $i=j$ in $f_{ab}$. 

Let us now suppose that we have an $\CN=2$ matrix integral associated to a quiver with $r$ nodes, where the CS levels are given again by (\ref{CSlevels}). 
We use the Cauchy identity (\ref{cauchy}) for each node of the quiver. It is then easy to see that the one-body potential associated to the node is given by 
\be
\label{one-body2}
U_a(\lambda)=-{\ri n_a k \over 4 \pi}\lambda^2 -N_{f_a} \ell\left( \widetilde \Delta_{f_a} + \ri {\lambda \over 2 \pi}\right) -N_{\tilde f_a} \ell\left( \widetilde \Delta_{\tilde f_a} - \ri {\lambda \over 2 \pi }\right), 
\ee
where $N_{f_a}$ ($N_{\tilde f_a}$) is the number of (anti) fundamentals in the $a$-th node. If there are fields in the bifundamental connecting the edges $a$ and $b$, the function $f_{ab}$ is given by
\be
f_{ab}(y)= -\ell\left( \widetilde \Delta_{(a,b)} + \ri {y \over 2 \pi} \right)-  \ell\left( \widetilde \Delta_{(b,a)} - \ri {y \over 2 \pi} \right)-\log \left( 2 \cosh{y \over 2} \right),
\ee
so that the function introduced in (\ref{gfx}) is 
\be
g_{ab}(y) = \exp\left[ \ell\left( \widetilde \Delta_{(a,b)} + \ri {y \over 2 \pi} \right)+  \ell\left( \widetilde \Delta_{(b,a)} - \ri {y \over 2 \pi} \right) \right]. 
\ee
Fields in the adjoint representation contribute to the function that we have denoted by $W_b(\mu_i -\mu_j)$, which is then given by 
\be
\sum_{i,j}W_b(\mu_i -\mu_j)=N \ell\left( \widetilde \Delta_{a} \right)+ \sum_{i<j}\left[ \ell\left(  \widetilde \Delta_{a} + \ri {\mu_i -\mu_j \over 2 \pi} \right) + 
\ell\left(  \widetilde \Delta_{a} - \ri {\mu_i -\mu_j \over 2 \pi} \right)  \right].
\ee

As in \cite{mp}, we identify $\rho_{ab}$ as a canonical density matrix, defining a Hamiltonian for a Fermi gas. To write this Hamiltonian in a 
more explicit form we use the Wigner transform. We recall that the Wigner transform of an operator $\hat A$ is given by (see \cite{hillery,cz} for a detailed exposition of phase-space quantization)
\be
\label{wignert}
A_{\rm W}(q,p)=\int \rd q' \left\langle q-{q'\over 2}\right|\hat A \left| q+{q'\over 2}\right\rangle \re^{\ri p q'/\hbar}.
\ee
The Wigner transform of a product is given by the $\star$-product of their Wigner transforms,
\be
\label{starprod}
\left(\hat A \hat B\right)_{\rm W}=A_{\rm W}\star B_W,
\ee
where the star operator is given by
\be
 \star=\exp\left[ {\ri \hbar \over 2} \left( {\overleftarrow{\partial}}_q {\overrightarrow{\partial}}_p  - {\overleftarrow{\partial}}_p {\overrightarrow{\partial}}_q\right) \right], 
 \ee
and 
\be
\label{planck}
\hbar = 2 \pi k.
\ee
It is convenient at this point to rescale the variables as $\lambda\rightarrow \lambda/k$. The Wigner transform of the canonical density matrix is then given by, 
\be
\label{rhowigner}
\rho_{ab}^{\rm W}= \re^{-\sum_i U_a (q_i/k)} \star F_{ab} (\{q_i\}, \{p_i\}) \star \re^{-\sum_{i,j} W_b\left({q_i-q_j\over  k}\right) }, 
\ee
where
\be
\label{fourier}
F_{ab}(\{q_i\}, \{p_i\}) =\int \rd y_1 \cdots \rd y_N \, \re^{\sum_{i=1}^N {\ri p_i y_i \over \hbar}} \prod_{i\not =j} \exp\left\{ -f_{ab}\left({q_i -q_j \over k}+{1\over 2} {y_i+ y_j\over k} \right) \right\} 
\prod_{i=1}^N g_{ab}(y_i/k). 
\ee
This can be also written as
 \be
 \ba
 \label{Fab}
 F_{ab}(\{q_i\}, \{p_i\}) &=\int \prod_{i=1}^N \rd y_i \,  g_{ab} (y_i/k) \re^{ \ri p_i y_i \over \hbar}
 \prod_{i\not =j}  \re^{-{\ri \hbar \over 2} \left(  {\overleftarrow{\partial}}_{p_i} {\overrightarrow{\partial}}_{q_{ij}}  +{\overleftarrow{\partial}}_{p_j} {\overrightarrow{\partial}}_{q_{ij}}  \right)} \exp\left\{ -f_{ab}\left({q_i -q_j \over k} \right) \right\}\\
 &=\prod_{i=1}^N \widehat g_{ab}(p_i)  \prod_{i\not =j}  \re^{-{\ri \hbar \over 2} \left(  {\overleftarrow{\partial}}_{p_i} {\overrightarrow{\partial}}_{q_{ij}}  +{\overleftarrow{\partial}}_{p_j} {\overrightarrow{\partial}}_{q_{ij}}  \right)} \exp\left\{ -f_{ab}\left({q_i -q_j \over k} \right) \right\},
 \ea
\ee
where
\be
\label{fourierg}
 \widehat g_{ab}(p)=\int \rd y \,  g_{ab}(y/k) \re^{ {\ri p y \over \hbar}}.
 \ee
The above construction defines the $N$-body Hamiltonian associated to $\rho_{ab}$ as 
\be
\rho_{ab}^{\rm W} = \re_{\star}^{-H^{ab}_{\rm W}}.
\ee
The total density matrix associated to the quiver is simply obtained by taking the product of the $\rho_{ab}$ over all edges. This is based on the 
identiy
\be
\ba
&\int \rd \lambda^{(1)} \cdots \rd \lambda^{(r)} K_1 \left( \lambda^{(1)}_1, \cdots, \lambda^{(1)}_N; \lambda^{(2)}_1, \cdots, \lambda^{(2)}_N\right)\cdots 
K_r\left( \lambda^{(r)}_1, \cdots, \lambda^{(r)}_N; \lambda^{(1)}_1, \cdots, \lambda^{(1)}_N\right)\\ &
=\int \rd \lambda^{(1)} \left\{ \lambda^{(1)}_1, \cdots, \lambda^{(1)}_N \right| \hat \rho_{K_1} \cdots \hat \rho_{K_r} \left| \lambda^{(1)}_1, \cdots,  \lambda^{(1)}_1\right\},
\ea
\ee
which just follows from using the resolution of the identity in $\CF_N$, 
\be
 \int \rd \lambda  \left| \lambda_1, \cdots,  \lambda_1\right\}\left\{ \lambda_1, \cdots, \lambda_N \right| ={\bf 1},
\ee
$r-1$ times. The corresponding, total Hamiltonian is simply obtained by taking the star product of the Wigner transforms $\rho_{ab}^{\rm W}$. For example, if we have a circular quiver, we obtain 
\be
\re_\star^{-H_{\rm W}}=\re_\star^{-H^{12}_{\rm W} } \star \, \re_\star^{-H^{23}_{\rm W}} \star \cdots \star \, \re_\star^{-H^{r-1 r}_{\rm W}} \star \,  \re_\star^{-H^{r1}_{\rm W}}.
\ee

In theories with $\CN\ge 3$ supersymmetry and canonical anomalous dimensions $\widetilde \Delta =1/2$, the resulting Fermi gas is a non-interacting one, as shown in \cite{mp}. We then 
obtain a one-body Hamiltonian whose kinetic term, at leading order in $\hbar$, is given by 
\be
-\log \widehat g_{ab} (p)= \log\left(2 \cosh {p \over 2}\right). 
\ee
For a general quiver, the Hamiltonian one obtains from the above procedure is quite complicated. At leading order in $\hbar$, we have a one-body Hamiltonian whose potential term is 
given by (\ref{one-body2}), summed over all nodes, and a kinetic term given by $-\log \,\widehat g_{ab}(p)$, summed over all edges. The functions $f_{ab}$, $W_b$ lead to two-body interactions. The function $W_b$ leads to a standard interaction between the fermions, while the function $f_{ab}$ leads to a more complicated, non-standard interaction between fermions which is velocity-dependent (since its Wigner transform involves both the positions $q_i$ and the momenta $p_i$). In addition, when considering $\hbar$ corrections, one will get $k$-body interactions from the commutators of these one-body and two-body interactions. As we will see in the next section, an important simplification occurs in the theories with one single node. Before looking at this case, we will analyze the fermion system at large $N$ and see how the functional theory developed in \cite{hklebanov,jkps} appears as a mean-field theory for this Fermi system.

\subsection{Large $N$ limit and Thomas--Fermi approximation}

The calculation of physical quantities in an interacting Fermi system and in the presence of an external potential is a non-trivial problem, and one is usually led to approximation schemes. One such scheme, which in some cases becomes exact where the number of particles $N$ is very large, is the Hartree approximation and its semiclassical limit, the Thomas--Fermi approximation. Let us quickly review some ingredients of this approximation, in the simple case in which we have a one-body Hamiltonian of the form 
\be
h(q,p)= T(p) + U(q)
\ee
and a two-body interaction with potential $V(q,q')$. The Hartree approximation can be regarded as a saddle-point evaluation of the many-body path integral (see for example \cite{no,kunz}). This saddle is characterized by a mean-field density $\rho(q)$ and an effective one-body potential
\be
U_{\rm eff}(q)=U(q)+ \int \rd q' \, V(q, q') \rho(q').
\ee
The second term in the r.h.s. is the Hartree (or direct) term. The density $\rho(q)$ is detemined in a self-consistent fashion by the Hartree equation (at finite temperature) 
\be
\label{hartree}
\rho(q) = n_{h_{\rm eff}}(q,q), 
\ee
where
\be
n_{H}(q,q')=\left\langle q \left|{1\over \re^{\beta (\hat H-\mu)}+1} \right|q'\right\rangle
\ee
is the matrix element of the average number operator, $\mu$ is the chemical potential, and the effective Hamiltonian $h_{\rm eff}$ is
\be
h_{\rm eff} (q,p)=T(p)+ U_{\rm eff}(q). 
\ee
The density satisfies the normalization condition
\be
\label{norma}
\int \rd q \, \rho(q)=N(\mu).
\ee
which gives the relation between $N$, the number of particles, and the chemical potential $\mu$. From this one can compute the grand potential in the Hartree approximation, 
\be
\label{grand-pot}
{\partial J \over \partial \mu}=\int \rd q \, \rho(q),
\ee
and therefore all the thermodynamic properties of the system. 

In general, the Hartree equation (\ref{hartree}) is not easy to solve. However, one can do further approximations. First, we note that the diagonal matrix element 
can be evaluated in terms of Wigner transforms, 
\be
n_{H}(q,q)=\int {\rd p \over 2 \pi \hbar} n_H^{\rm W}(q,p). 
\ee
On the other hand, in the semiclassical limit, we can use the classical Hamiltonian, 
\be
n_H^{\rm W}(q,p)\approx {1\over \re^{\beta(H(q,p)-\mu)}+1},
\ee
and in this limit (\ref{hartree}) reads 
\be
\label{tfT}
\rho(q)= \int {\rd p \over 2 \pi \hbar}  \left[ \exp\left( \beta \left( T(p)+ U_{\rm eff}(q)  -\mu\right) \right)+1\right]^{-1}.
\ee
 This is the Thomas--Fermi equation at finite temperature, see for example \cite{tomi}. A further approximation involves going to zero temperature. Then, the semiclassical occupation number is determined by 
 \be
 n_H^{\rm W}(q,p)\approx \Theta \left(\mu-H(q,p) \right). 
 \ee
The ``classical" Fermi surface,
\be
H(q,p)=\mu, 
\ee
defines implicitly the so-called local Fermi momentum\footnote{In this part we assume that the Hamiltonian is symmetric in momentum.} $p_F(q, \mu)$. The density $\rho(q)$, in the semiclassical limit and at zero temperature, 
is called the Thomas--Fermi density. In the zero-temperature Thomas--Fermi limit, (\ref{hartree}) can be regarded as an extremization condition for a density functional, the Thomas--Fermi functional, 
\be
E_{\rm TF}[ \rho]=t_{\rm TF}(q) + \int \rd q \, \rho(q) U(q) + {1\over 2} \int \rd q \rd q' \rho(q) V(q-q') \rho(q') -\mu \left(\int \rd q \, \rho(q)-N\right) ,
\ee
where $t_{\rm TF}(q)$ is the kinetic energy functional, 
\be
 t_{\rm TF}(q)=\int {\rd p \over 2 \pi \hbar} T(p) \theta\left(\mu-h_{\rm eff}(q,p)\right). 
\ee
Notice that, in one dimension, the Thomas--Fermi density $\rho(q)$ is always proportional to the local Fermi momentum, 
\be
\label{Fermirho}
\rho(q) ={1\over \pi \hbar} p_F(q, \mu), 
\ee
where in the r.h.s. we have taken the positive solution for the local Fermi momentum. 

Of course, for an ``atom," i.e. a three-dimensional Fermi gas in an external, attractive Coulomb potential, and with mutual Coulomb repulsion, the above formalism leads to the standard Thomas--Fermi approximation in atomic physics. This approximation become exact as $N \rightarrow \infty$ \cite{lieb}. As an even simpler 
example of the Thomas--Fermi formulation, we can consider the non-interacting Fermi gas which appears in ABJM theory \cite{mp}. 
In ABJM theory, the Hamiltonian is, at leading order in $\hbar$, 
\be
H(q,p) =\log \left(2 \cosh {p\over 2}\right) + \log \left(2 \cosh {q\over 2}\right)\approx {|p| \over 2} +{|q| \over 2}.
\ee
The Thomas--Fermi kinetic functional is 
\be
t_{\rm TF}(q)=\int {\rd p \over 2 \pi \hbar} {|p|\over 2}\theta(\mu-H(q,p)) ={\pi \hbar \over 4} \rho^2(q). 
\ee
Therefore, the Thomas--Fermi functional is simply (up to the constant $\mu N$)
\be
\label{funct-abjm}
E_{\rm TF}[\rho]= \int \rd q\, \left[ {\pi \hbar \over 4} \rho^2(q) +{|q|\over 2} \rho(q)-\mu \rho(q)\right].
\ee
Extremizing this functional leads to
\be
\label{abjm-density}
\rho(q)={2 \mu - |q| \over \pi \hbar}, 
\ee
which is just (\ref{Fermirho}) for the Fermi momentum obtained from the Hamiltonian. 
The constraint of having $N$ particles gives
\be
\mu={\sqrt{\pi \hbar N} \over 2}
\ee
and the evaluation of this functional on the above solution reproduces the well-known result \cite{dmp}, 
\be
E_{\rm TF}[\rho]={\sqrt{2} \over 3} \pi k^{1/2} N^{3/2}.
\ee
The general Hartree/Thomas--Fermi approximation is supposed to be exact, at large $N$, if the system becomes very dense in that limit (see \cite{wittenbaryon} for some useful comments on this issue). We see from (\ref{abjm-density}) that the size of the Fermi gas scales like $N^{1/2}$, therefore the density grows like $N^{1/2}$ at large $N$, and indeed we are dealing with a dense system. This behavior is in fact typical of all the systems considered in \cite{mp} and in this paper, so the Thomas--Fermi approximation is appropriate.  

The functional (\ref{funct-abjm}) is very much like the one in \cite{hklebanov} 
after the $y$ variable has been integrated out. The factor $\rho^2$, which in \cite{hklebanov} arises from the interaction between eigenvalues, is due here 
to the kinetic term in the Thomas--Fermi functional. We will now show that a mean-field treatment of the general interacting Fermi gas considered above leads to a functional of the Thomas--Fermi density $\rho$ which is identical to the one obtained in \cite{jkps,ms,k1}. 

First, let us note that there is a freedom in the choice of $\rho$ in (\ref{K-rho-relation}) for a given $K$. In principle one can simply choose 
\begin{equation}
\rho_{ab}(\lambda_1, \cdots, \lambda_N; \mu_1, \cdots, \mu_N)=K_{ab}(\lambda_1, \cdots, \lambda_N; \mu_1, \cdots, \mu_N).
\end{equation} 
Let us separate in this kernel the one-loop and the classical contribution due to the CS action:
\begin{equation}
 K_{ab}(\{\lambda_i\},\{\mu_i\})= \re^{\sum \frac{k_a \mu_i^2 }{4\pi}}\,\tilde{K}_{ab}(\{\lambda_i\},\{\mu_i\}).
\end{equation} 
The Wigner transform of the one-loop part is given by
\begin{equation}
 \tilde{K}_{ab}^\wigner(\{q_i\},\{p_i\})=\int \prod_i \rd y_i\, \re^{\ri\sum_i p_iy_i/\hbar}\,
 \tilde{K}_{ab}\left(\left\{\frac{q_i}{k}+\frac{y_i}{2k}\right\},\left\{\frac{q_i}{k}-\frac{y_i}{2k}\right\}\right).
 \label{Kab-WT}
\end{equation} 
According to the general principles of the Thomas-Fermi approximation, one can suppose that the fermions are distributed uniformly with the density $1/(2\pi\hbar)$ in the domain of the phase space bounded by the upper and lower Fermi momenta $p_F^+(q)$ and $p_F^-(q)$. Then the density of the distribution in the coordinate space is given by
\begin{equation}
 \rho(q)=\frac{p_F^+(q)-p_F^-(q)}{2\pi\hbar}.\label{rho-p-fermi}
\end{equation} 
Let us note that in (\ref{Kab-WT}) $y_i$ plays the role of the difference of the eigenvalues associated to adjacent nodes. Suppose that the main contribution is given by imaginary $y_i$ (after deformation of the contour of the integration). Then, by taking a continuous limit, 
\be
y_i\rightarrow \ri y_{ab}(q),
\ee
the kernel $\tilde{K}_{ab}$ can be computed by repeating the calculations done in Appendix A of \cite{jkps} (see also \cite{ghp1,ghp2,hklebanov}). It has the following form 
(for the sake of simplicity we consider the case without fundamental matter multiplets):
\begin{equation}
\tilde{K}_{ab}\left(\left\{\frac{q_i}{k}+\frac{y_i}{2k}\right\},\left\{\frac{q_i}{k}-\frac{y_i}{2k}\right\}\right)
 \approx \exp\left\{\int \rd q\rho^2(q)\phi_{ab}(y_{ab}(q))\right\},
\end{equation}
where 
\begin{equation}
 \phi_{ab}(y)=kf_{\Delta_{(a,b)}}\left(\frac{y}{2\pi k}\right)+kf_{\Delta_{(b,a)}}\left(-\frac{y}{2\pi k}\right)+kf_{\Delta_{a}}(0),
\end{equation} 
\begin{equation}
 f_\Delta(z)=\frac{2\pi^2}{3}(z+\Delta)(z+\Delta-1)(z+\Delta-2),\qquad 0\leq \Delta +z\leq 2.
\end{equation} 
Also
\begin{equation}
 \frac{\ri}{\hbar}\sum_i p_iy_i \approx -\frac{1}{2\pi\hbar^2}\int \rd q\int\limits_{p_F^-(q)}^{p_F^+(q)} \rd p\, p\,y_{ab}(q)=-\frac{1}{\hbar}\int \rd q\,\rho(q)\bar{p}(q)\,y_{ab}(q),
\end{equation} 
where
\begin{equation}
 \bar{p}(q)\equiv \frac{p_F^+(q)+p_F^-(q)}{2}.\label{pbar-p-fermi}
\end{equation} 
The Wigner transform of the total density matrix is given by
\begin{multline}
 \rho^\wigner(\{q_i\},\{p_i\})=
 \starprod_{a=1}^r K_{a,a+1}^\wigner(\{q_i\},\{p_i\})=
 \starprod_{a=1}^r \re^{\sum_i \frac{n_{a} q_i^2 }{2\hbar}}\star \tilde{K}_{a,a+1}^\wigner(\{q_i\},\{p_i\}) =\\
= \starprod_{a=1}^r \tilde{K}_{a,a+1}^\wigner(\{q_i\},\{p_i-\sum_{b=1}^an_b\, q_i\}).
\end{multline}
Then
\begin{equation}
 \rho^\wigner(\{q_i\},\{p_i\})
\approx \int \mathcal{D}
y \exp\sum_a\int\rd q\rho(q)\left\{\frac{Q_aq-\bar{p}(q)}{\hbar}\cdot y_{a,a+1}(q)
 +\rho(q)\phi_{a,a+1}(y_{a,a+1}(q))
 \right\},
\end{equation}
where $Q_a=\sum_{b=1}^a n_b$ and we replaced the $\star$-product by the ordinary one. Using the saddle point approximation for the integration over $y_{a,a+1}$ we arrive at the following expression for the Thomas-Fermi functional:
\begin{multline}
 E_\text{TF}[p_F^+,p_F^-]=\min_{y_{a,a+1}}\sum_a\int\rd q\rho(q)\left\{\frac{Q_aq-\bar{p}(q)}{\hbar}\cdot y_{a,a+1}(q)
 -\rho(q)\phi_{a,a+1}(y_{a,a+1}(q))
 \right\}\\
 -\mu \left(\int \rd q \, \rho(q)-N\right),
\end{multline}
 where $\rho(q)$ and $\bar{p}(q)$ are related to the Fermi momenta $p_F^+(q)$ and $p_F^-(q)$ through (\ref{rho-p-fermi}) and (\ref{pbar-p-fermi}). The thermodynamic limit of the free energy is given by
\begin{multline}
 -F=\min_{p_F^+,p_F^-,\mu} E_\text{TF}[p_F^+,p_F^-]=\\
=\min_{\rho,\bar{p},y_{a,a+1},\mu}\left[
\sum_a\int\rd q\rho(q)\left\{\frac{Q_aq-\bar{p}(q)}{\hbar}\cdot y_{a,a+1}(q)
 -\rho(q)\phi_{a,a+1}(y_{a,a+1}(q))
 \right\}
\right.\\
\left.
-\mu \left(\int \rd q \, \rho(q)-N\right)
\right].
\end{multline} 
This minimization prescription is the same as the one that was first introduced in \cite{hklebanov}. The function $\bar{p}$ can be considered as the Lagrange multiplier for the condition $\sum_a y_{a,a+1}=0$ (cf. \cite{ghp1}).

\sectiono{Theories with one single node: flavored theories}

\subsection{The Thomas--Fermi approximation}

We consider now the flavored theories with one single node studied in \cite{benini,jafferis} and whose matrix model was analyzed in section 4 of \cite{jkps}. These theories are much simpler from the point of view of the picture in terms of interacting fermions, since at leading order in the $\hbar$ expansion they have only position-dependent interactions. 

The theories under consideration have a single node with gauge group $U(N)$, and three sets of pairs of chiral superfields in the 
fundamental representation, 
\be
(q_j^{(i)}, \tilde q^{(i)}_j), \qquad i=1,2, 3, \quad j=1, 2, \cdots, n_i. 
\ee
Their anomalous dimensions are denoted by 
\be
\Delta_{q_i}, \, \Delta_{\tilde q_i}. 
\ee
There are also three adjoint chiral superfields 
\be
X_1, \, X_2, \, X_3. 
\ee
The anomalous dimensions of these fields are denoted by 
\be
\Delta_i, \qquad i=1,2,3. 
\ee
They satisfy the following constraints
\be
\label{nolong}
\sum_{i=1}^3 \Delta_i=2, 
\ee
and
\be
\label{delcons}
\Delta_{q_i} + \Delta_{\tilde q_i}+ \Delta_i=2. 
\ee
One has to add monopole operators \cite{jkps}, but as shown in that paper they do not contribute to the final answer for the free energy 
and we will not consider them. Their inclusion is however straightforward. We will introduce a parameter $k$ as follows, 
\be
n_i =k f_i, 
\ee
and we will formally regard it as the parameter for a semi-classical expansion in the Fermi gas, i.e. the Planck constant is given by the same relationship as (\ref{planck}). 
The matrix model computing the free energy on the three-sphere of this theory can be obtained with the rules reviewed in subsection \ref{matrix-rules}. Its integrand contains the factors 
\be
\label{flav-int}
\ba
&\prod_{i=1}^3 \prod_{m,n} \exp \left[ \ell\left( 1-\Delta_i + \ri {\lambda_m -\lambda_n \over 2 \pi}\right) \right] \\
& \times \prod_m \exp\left[ n_i \ell 
\left( 1-\Delta_{q_i} + \ri {\lambda_m  \over 2 \pi}\right)+ n_i \ell 
\left( 1-\Delta_{\tilde q_i} - \ri { \lambda_m  \over 2 \pi}\right) \right] 
\ea
\ee
due to the adjoint superfields and the fundamental superfields. In addition, we have the standard contribution 
\be
\prod_{m<n} \left[ 2 \sinh\left({\lambda_m -\lambda_n \over 2 } \right)\right]^2 
\ee
due to the $U(N)$ vector multiplet. We now introduce the quantity 
\be
\CQ(\{\lambda_i\})={\prod_{m<n} \left( 2 \sinh {\lambda_m -\lambda_n \over 2} \right)^2 \over
\prod_{m,n}  2 \cosh {\lambda_m -\lambda_n \over 2} }=\sum_{\sigma \in S_N} (-1)^{\epsilon(\sigma)} {1\over \prod_m 2 \cosh\left( {\lambda_m -\lambda_{\sigma(m)} 
\over 2} \right)}. 
\ee
The interaction between eigenvalues in the one-node matrix model can then be written as 
\be
\label{intone}
a(N) \CQ\left(\left\{\frac{q_i}{k}\right\}\right) \prod_{m<n} \re^{-V(q_m- q_n)},
\ee
where
\be
\lambda_m ={q_m \over k},
\ee
the potential is given by
\be
\label{potonode}
V(q)= -\sum_{i=1}^3 \left[ \ell\left( \widetilde \Delta_i + \ri {q \over 2k \pi}\right)+ \ell \left( \widetilde \Delta_i -\ri {q \over 2k \pi}\right) \right]- 2 \log \left( 2\cosh {q \over 2 k}\right)
\ee
and the prefactor in (\ref{intone}) is
\be
\label{prefactor}
a(N)=\exp \left\{ N  \left( \log 2 + \sum_i \ell \left(\widetilde \Delta_i \right) \right) \right\}.
\ee
It comes from the diagonal terms $m=n$ not included in (\ref{intone}).  

After using the Cauchy identity, the integrand of the matrix model can be written as 
\be
 \left\{ \lambda_1, \cdots, \lambda_N \right| \hat \rho \left| \lambda_1, 
\cdots, \lambda_N\right\},
\ee
where
\be
\rho (\lambda_1, \cdots, \lambda_N; \mu_1, \cdots, \mu_N)=\prod_{m=1}^N \re^{-U(k\lambda_m)} \prod_{m<n} \re^{-V(k\lambda_m- k\lambda_n)} \prod_{m=1}^N t(\lambda_m-\mu_m), 
\ee
$t(y)$ is given in (\ref{tx}), and the one-body potential can be read from (\ref{flav-int}) and it reads, 
\be
U(q)=-\sum_{i=1}^3n_i  \left[ \ell 
\left( 1-\Delta_{q_i} + \ri {q \over 2 \pi k}\right)+ \ell 
\left( 1-\Delta_{\tilde q_i} - \ri { q  \over 2 \pi k}\right) \right]. 
\ee
The Wigner transform of $\rho$ is given by 
\be 
\rho_\text{W}(\{q_i\}, \{p_i\})= \prod_{m=1}^N \re^{-{1\over 2} U(q_m)} \prod_{m<n} \re^{-{1\over 2} V(q_m- q_n)}\star \prod_{m=1}^N {1\over 2 \cosh\left( {p_m \over 2} \right)} \star \prod_{m=1}^N \re^{-{1\over 2} U(q_m)} \prod_{m<n} \re^{-{1\over 2} V(q_m-q_n)}
\ee
Up to corrections coming from higher orders in the star product, this is a gas of $N$ particles with kinetic term 
\be
T(p)= \log \left( 2 \cosh {p \over 2} \right), 
\ee
as in \cite{mp}, and a one-particle potential $U(q)$. At large $|q|$ this potential becomes
\be
\label{polyU}
U(q)= \gamma {|q| \over 2} +\cdots, 
\ee
where
\be
\gamma=\sum_{i=1}^3 f_i\Delta_i.
\ee
The interaction potential between the particles in this gas is 
\be
\label{twobody}
V(q)=\sum_{i=1}^3 v_{\Delta_i} \left( {q\over k}\right) -2 v_{1/2} \left( {q\over k}\right), 
\ee
since, due to (\ref{nolong}), the long range potential proportional to $|q|$ cancels. The resulting potential is a repulsive, short-range potential. Of course, on top of the one-body and the two-body potential, the star product leads to an infinite series of ``quantum" corrections involving 
$s$-body potentials, which are in addition velocity-dependent (they depend on both $q_i$ and $p_i$). The quantum Hamiltonian $H_\wigner$ can be computed by using the Baker-Campbell-Hausdorff formula. One finds:
\begin{multline}
 H_\wigner(\{q_i\}, \{p_i\})=\sum_nU(q_n)+\sum_nT(p_n)+\frac{1}{2}\sum_{m\neq n}V(q_n-q_m)\\
-\frac{\hbar^2}{12}\sum_n\left(T'(p_n)\right)^2U''(q_n)-\frac{\hbar^2}{12}\sum_{n\neq m}T'(p_n)\left(T'(p_n)-T'(p_m)\right)V''(q_n-q_m)\\
+\frac{\hbar^2}{24}\sum_nT''(p_n)\left(U'(q_n)+\sum_{m\neq n}V'(q_n-q_m)\right)^2+\mathcal{O}(\hbar^4).
\end{multline}

We will now analyze the resulting Fermi gas after doing the following approximations: 
\begin{enumerate} 

\item We neglect the quantum corrections to the Hamiltonian. 

\item We treat the two-body interaction in the Hartree/Thomas--Fermi approximation reviewed above. 

\item We take the limit of zero temperature. 

\item We take the polygonal limit of the one-particle Hamiltonian, corresponding to large $|q|$ and $|p|$. 

\end{enumerate}

As we will see, we will recover in this way the solution in \cite{jkps} by considering the so-called Hartree approximation to the interacting problem 
and doing moreover the following approximations:

In the polygonal limit, the kinetic term is simply given by 
\be
 T(p) \approx {|p| \over 2}
 \ee
and the zero-temperature Thomas--Fermi equation reads in this case 
\be
\rho(q) =\int {\rd p \over 2 \pi \hbar} \Theta\left( \mu -{|p| \over 2} - U(q) -\int \rd q' \, V(q-q') \rho(q') \right). 
\ee
In one dimension, the Thomas--Fermi density is just given by the local Fermi momentum, 
\be
\rho(q) ={p_F(q) \over \pi \hbar}, 
\ee
where $p_F(q)$ solves the equation
\be
\label{localfermi}
p_F(q)= 2 \mu - 2 U(q) - 2 \int \rd q' \, V(q-q') \rho(q'). 
\ee
In the polygonal limit, in which we neglect exponential corrections to the functions appearing here, we can 
use (\ref{approxker}) to write
 \be
\int \rd q' \, V(q-q') \rho(q') \approx  \left( \int \rd q' \, V(q')\right) \rho(q). 
\ee
Using also (\ref{vint}) and (\ref{nolong}) we find, 
\be
\int \rd q' \, V(q-q') \rho(q') \approx \pi^2 k \left( \delta - 1\right) \rho, 
\ee
where
\be
\delta= 4 \Delta_1 \Delta_2 \Delta_3. 
\ee
After all these approximations, the solution to the Thomas--Fermi equation is simply 
\be
\rho(q)= {1\over  \pi^2 k \delta} \left( \mu -U(q)\right)
\ee
and we can use the polygonal approximation (\ref{polyU}) for $U(q)$. 
The effective potential is 
\be
U_{\rm eff}(q)=\left( 1- {1\over  \delta} \right) \mu+{1\over  \delta} U(q). 
\ee

\begin{figure}
\begin{center}
\includegraphics[scale=.4]{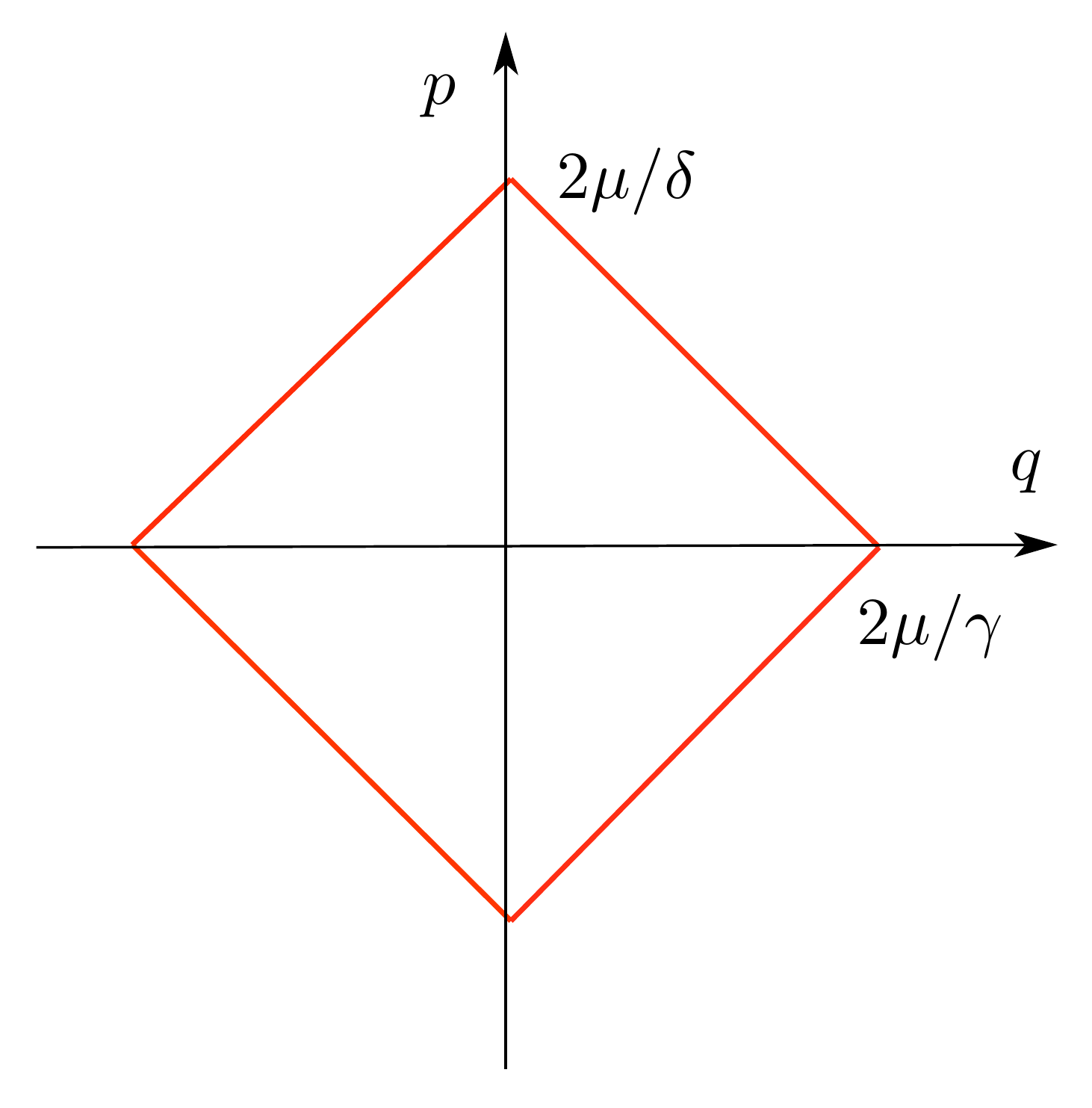}
\end{center}
\caption{The Fermi surface for the interacting Fermi gas associated to flavored one-node theories, in the Thomas--Fermi approximation.} 
\label{fig-fermis}
\end{figure}

In the Hartree approximation, we can think about the interacting problem as a one-body problem with the effective potential above. The Fermi surface is then determined by 
\be
{|p| \over 2} + U_{\rm eff}(q)=\mu, 
\ee
or equivalently
\be
{|p| \over 2} + {\gamma \over \delta}  {|q| \over 2} = {\mu \over \delta},
\ee
and we have depicted it in \figref{fig-fermis}. The area of this Fermi surface, measured in units of $2 \pi \hbar$, equals the number of particles of the gas, 
\be
N={ {\rm vol}(\mu) \over 2 \pi \hbar}= {1\over \delta \gamma} { 2 \mu^2 \over \pi^2 k}. 
\ee
This is of course equivalent to the normalization condition (\ref{norma}). The grand potential is, at leading order, 
\be
\label{leadinggp}
J(\mu)\approx {1\over \delta \gamma} {2 \mu^3 \over 3 \pi^2 k}. 
\ee
After an inverse Legendre transform, 
\be
\label{freethermo}
F(N) =J(\mu(N)) - N \mu(N),
\ee
we immediately find
\be
\label{leadingfn}
F(N)\approx -{ {\sqrt{2}} \pi \over 3} N^{3/2} k^{1/2} {\sqrt{ \delta \gamma}} =-{ 2 {\sqrt{2}} \pi \over 3} N^{3/2}  {\sqrt{ \Delta_1 \Delta_2 \Delta_3  \left(\sum_{i=1}^3 n_i \Delta_i \right)}}
\ee
which is exactly the result of \cite{jkps}. 

Notice that, in the mean-field picture coming from the Hartree/Thomas--Fermi approximation, we have $N$ fermions in the presence of an external, linear confining potential, and they fill out an interval of lenght $\sim \mu \sim N^{1/2}$, just like in ABJM theory. The effect of the short-range interaction is to modify the precise numerical value of the parameters involved in the solution, without changing the qualitative picture of the Fermi droplet. Since we have a very dense system at large $N$, we expect the Thomas--Fermi approximation to be exact in the large $N$ limit. Moreover, since the support of the Thomas--Fermi density grows like $N^{1/2}$, we are justified in taking the polygonal approximation, since corrections to this approximation will be subleading, as in \cite{mp} (as we will see in a moment, in the presence of a long--range attractive potential, the support of the Thomas--Fermi density will be of order $\CO(1)$, and the polygonal approximation breaks down.)

We can then interpret the functional obtained in \cite{jkps} as the Thomas--Fermi approximation to an interacting Fermi gas, leading to the correct result in the large $N$ limit. In particular, (\ref{leadinggp}) is the correct grand potential at leading order in $\mu$. 

\subsection{Corrections to the Thomas--Fermi approximation}

One advantage of the Fermi picture, already emphasized in \cite{mp}, is that one has in principle a systematic way of improving the leading large $N$ approximation. In the non-interacting case considered in \cite{mp}, this made possible to determine an infinite number of subleading $1/N$ corrections. The interacting case is much more difficult, but we expect the following corrections to the above result for the grand potential (\ref{leadinggp}): 

\begin{enumerate} 

\item As we already pointed out, without leaving the Hartree approximation, there will be corrections coming from finite $T$ effects and from deviations from the polygonal limit. These corrections already occur in the context of the ideal Fermi gas analyzed in \cite{mp}, and there are clearly present in this example as well. In order to calculate these corrections quantitatively we have to solve the Thomas--Fermi equation for $\rho(q)$ beyond the polygonal limit approximation. 

\item One has to take into account the quantum corrections to the Hamiltonian. There are corrections to the one-body Hamiltonian, as in \cite{mp}, as well as to the two-body Hamiltonian. In addition, the quantum corrections will lead to $s$-body interactions, for all $s\ge 3$, which should be also taken into account. Furthermore, since we are using the Thomas--Fermi approximation, one should also expect  
corrections to the semiclassical approximation, of the Wigner--Kirkwood type. 

\item As it is well-known in many-body theory, the Hartree approximation is the starting point for a (resummed) diagrammatic expansion. Even if we restrict ourselves to the two-body interaction (\ref{twobody}), there will be corrections coming from exchange and correlation effects. 

\end{enumerate}

We expect on general grounds that these corrections will lead to sub-leading terms in $\mu$ in the grand potential (\ref{leadinggp}). We don't have a systematic argument for this, but a detailed examination of many correction terms suggests that this is the case. More precisely, the next-to-leading correction to (\ref{leadinggp}) comes for the exchange correction to Hartree theory, and goes like $\mu^2$. However, the resulting correction to the partition function {\it cancels} against the prefactor (\ref{prefactor}). Let us see this in detail. 

\begin{figure}
\begin{center}
\includegraphics[scale=.5]{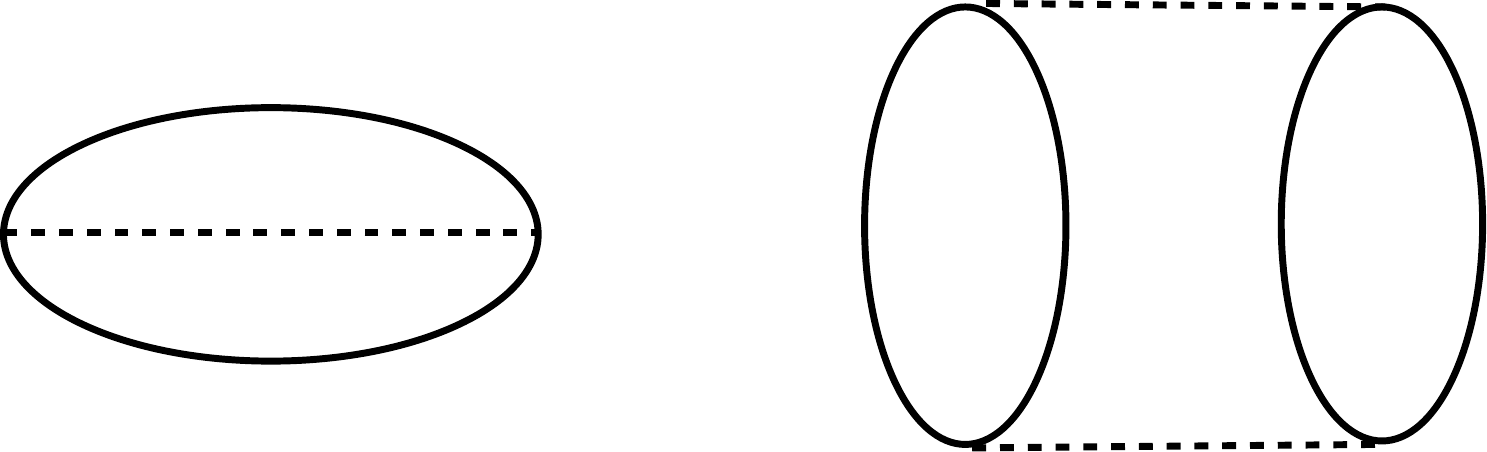}
\end{center}
\caption{The exchange correction due to the two-body interaction (\ref{twobody}) (left) and the first ring diagram contributing to correlation effects (right).} 
\label{fig-diagrams}
\end{figure}

The exchange correction to the Hartree approximation can be represented diagramatically as in \figref{fig-diagrams}, left (see for example \cite{no}). Its contribution to the grand potential is given by 
\be
J_{\rm ex}(\mu)={1\over 2} \int \rd q \, \rd q' \, n_{h_{\rm eff}} (q',q) V(q-q') n_{h_{\rm eff}} (q,q'). 
\ee
We will evaluate this correction in the same approximation scheme that we used before, i.e. we will use a semiclassical calculation at zero temperature, and in the polygonal limit (improving this approximation leads to even more subleading corrections). The two-point function of the occupation number operator appearing in this expression can be computed in terms of Wigner transforms as
\be
n_{h_{\rm eff}} (q',q)=\int {\rd p \over 2 \pi \hbar} \re^{\ri p r \over \hbar} n_{h_{\rm eff}}^{\rm W}(R, p), 
\ee
where 
\be
R={q+q'\over 2}, \qquad r=q-q'.
\ee
In the semiclassical and zero temperature approximation we have 
\be
n_{H_{\rm eff}} (q',q)\approx 
\int {\rd p \over 2 \pi \hbar} \re^{\ri pr \over \hbar} \Theta\left(\mu- h_{\rm eff}(p,R) \right)={1\over \pi} {\sin \left( p_F(R)r /\hbar\right) 
\over r},
\ee
where $p_F(q)$ is the local Fermi momentum for the effective Hamiltonian. 
Therefore, the exchange term reads
\be
J_{\rm ex}(\mu)\approx {1\over 2 \pi^2 } \int \rd R \, \rd r \, \left( {\sin \left( p_F(R)r /\hbar\right) 
\over r}\right)^2 V(r).
\ee
Let us estimate this for large $\mu$. In the polygonal approximation, $p_F$ is given by 
\be
p_F(q)= {2 \mu \over  \delta}-\frac{\gamma}{\delta} |q|
\ee
and it scales as $\mu$. We introduce now the rescaled coordinates $\xi, \zeta$ as
\be
r=\xi/\mu,\qquad R=\mu \zeta
\ee
so that the exchange integral reads
\be
J_{\rm ex}(\mu)\approx {\mu^2 \over 2 \pi^2 } \int \rd \zeta \, \rd \xi \, \left( {\sin \left( \hat p_F (\zeta) \xi/\hbar\right) 
\over \xi}\right)^2 V(\xi/\mu),
\ee
 where
 \be
p_F(R) = \mu \hat p_F (\zeta), \qquad \hat p_F (\zeta)= {2\over \delta}-\frac{\gamma}{\delta} |\zeta|. 
\ee
For $\mu$ large, we have then 
\be
\ba
J_{\rm ex}(\mu) \approx {\mu^2 \over 2 \pi^2 } V(0) \int \rd \zeta \, \rd \xi \, \left( {\sin \left( \hat p_F (\zeta) \xi/\hbar\right) 
\over \xi}\right)^2 ={\mu^2 \over 2 \pi \hbar} V(0) \int \rd \zeta \hat p_F(\zeta)={1\over 2} V(0) N,
\ea
\ee
where we used the normalization condition (\ref{norma}). However, we know from (\ref{potonode}) that
\be
V(0)=-2 \sum_{i=1}^3 \ell\left(\widetilde \Delta_i\right)-2 \log 2, 
\ee
therefore the leading, $\mu^2$ contribution of the exchange term cancels against the prefactor (\ref{prefactor}) in (\ref{intone}) (a similar cancelation occurs in mean-field many body theory, see section 7.2 of \cite{no}). 

We have not found other sources for $\mu^2$ corrections. It is easy to see that there are corrections of order $\CO(\mu)$ (these corrections are already present in the non-interacting 
case considered in \cite{mp}) and of order $\CO(1)$. For example, the first ring diagram showed in \figref{fig-diagrams} (right) can be evaluated in the semiclassical limit with the techniques of 
\cite{kunz}, and it can be seen to be of order $\CO(1)$. We then expect the grand potential to be of the form 
\be
J(\mu)={1\over \delta \gamma} {2 \mu^3 \over 3 \pi^2 k} + B \mu + A + \CO\left(\re^{-\mu}\right), 
\ee
where $B$, $A$ do not depend on $\mu$. By using the standard inversion formula 
\be
\label{muint}
Z(N) ={1\over 2 \pi \ri} \int \rd \mu \, \exp\left[J(\mu) - \mu N\right],
\ee
much exploited in \cite{mp}, we conclude that the partition function of the single-node theories is of the form 
\be
Z(N) =C^{-1/3}\re^{A}\Ai\left[C^{-1/3}(N-B)\right],
\end{equation} 
up to non-perturbative corrections in $N$. Here,
\be
C={2 \over  \pi^2 \delta \gamma k}.
\ee
This would confirm the conjecture made in \cite{mp} for this class of $\CN=2$ theories. It would be important to perform a systematic analysis of the possible corrections to $J(\mu)$ in order to verify our 
preliminary analysis here. It would be also interesting to see if one can calculate the coefficient $B$ by using the interacting Fermi gas picture. In principle, it is clear what one should do: among other things, one has to analyze the Hartree equation beyond the zero temperature and the polygonal approximation, and one has to take into account the corrections coming from many-body diagrams. 

\sectiono{Theories with one single node: long-range forces}

We will now consider a different class of theories with one single node, namely $\CN=2$, $U(N)$ Chern--Simons theory with $g$ adjoint multiplets. These theories were first studied in \cite{gy}, and the matrix model partition function was analyzed in \cite{minwalla,amariti} (for general $g$) and in \cite{niarchos,mn} (for $g=1$). Here, we will study these theories from the point of view of interacting fermions developed in this paper. This will allow us to re-derive and improve some of the results in \cite{minwalla}. In particular, we will see how the Thomas--Fermi equation of this model -- a single integral equation -- determines the $R$-charge of the adjoint multiplets in the large $N$ limit and at infinite 't Hooft coupling in a very efficient way (the method used in \cite{minwalla} was based on the numerical extrapolation of the saddle-point equations at finite $N$ and 't Hooft coupling). We also compute the large $N$ free energy of these models from the Thomas--Fermi equation. 

The partition function on the three-sphere of $\CN=2$, $U(N)$ Chern--Simons theory with $g$ adjoint multiplets is given by 
\be
\label{gad-mm}
Z=\int \prod_{i=1}^N \rd \lambda_i \re^{{\ri k \over 4 \pi} \lambda_i^2} \, \prod_{i<j} \left( 2 \sinh  {\lambda_i-\lambda_j  \over 2} \right)^2 \exp\left[g \sum_{i,j} \ell\left( 1-h + \ri {\lambda_i -\lambda_j \over 2 \pi} \right) \right]. 
\ee
This follows from the rules reviewed in section 2. We have denoted by $h=\Delta_{\rm Ad}$ the $R$-charge of the adjoint hypermultiplet. In principle, this partition function depends on four parameters: $N$, $k$, $g$ and $h$. However, the value of $h$ is determined, as a function of $N$, $k$, and $g$, by maximizing $|Z|$ \cite{jafferis}, so in fact there are only three free parameters.

 If we use the Fermi interpretation developed in section 2, we immediately find a long-range potential between the fermions of the form 
\be
\label{lr-pot}
\left( g (1-h) -1\right) |x|.
\ee
Clearly, the nature of the fermionic system will depend crucially on the sign of the coefficient. If this sign is positive, we have a {\it long-range attraction}, and if the sign is negative we have a {\it long-range repulsion}. We will analyze these two situations separately. In both cases, as we will see, the Thomas--Fermi distribution is supported on an interval whose length does not longer scale with $N^{1/2}$, as in the $\CN\ge 2$ theories considered above and in \cite{dmp,hklebanov,jkps,mp}. In the case where there is long-range attraction, the support of the distribution has a length of order $\CO(1)$. This means that, even at large $N$, we cannot use the polygonal approximation that we used in the case of a short-range potential, since the terms we would neglect in this approximation are as important as the terms that we would keep. Moreover, the quantum corrections to the Hamiltonian are also important in this case. 

The easiest way to incorporate all these corrections in the Thomas--Fermi equation is to take the standard large $N$ limit directly in the matrix integral (\ref{gad-mm}), in which the Thomas--Fermi distribution $\rho(x)$ is interpreted as a density of eigenvalues. Standard techniques lead to the integral equation
\begin{equation}
 \int \rd y \rho(y) K(x-y)=\mu-\frac{\alpha x^2}{4\pi},\qquad x\in\mathrm{supp}\,\rho,
\label{integral-eq}
\end{equation}
where 
\be
\alpha=-\ri k
\ee
and the kernel $K$ is given by
\begin{equation}
 K(x)=g V_{h}(x)-2\log 2\sinh \frac{|x|}{2},
\end{equation}
where
\begin{equation}
 V_h(x)=-\ell\left(1-h+\ri\frac{x}{2\pi}\right)-\ell\left(1-h-\ri\frac{x}{2\pi}\right).
\end{equation}
In what follows we suppose that $\alpha$ is real and positive to ensure that the eigenvalues are distributed along the real axis. The free energy for real values of $k$ can then be obtained by analytical continuation.

At large $x$, the kernel behaves as 
\begin{equation}
 K(x)= (g(1-h)-1)|x|+\CO(\re^{-|x|}),
\end{equation} 
which is the long-range potential (\ref{lr-pot}). At small $x$, it behaves as
\begin{equation}
 K(x)=-2\log|x|+gA_0+\sum_{k\geqslant 1}\left(A_{2k}-\frac{B_{2k}}{k}\right)\,\frac{x^{2k}}{(2k)!},
\end{equation} 
where $A_{2k}=V^{(2k)}_h(0)$ and $B_{2k}$ are Bernoulli numbers. As we mentioned before, the distribution of eigenvalues depends drastically on the sign of $(g(1-h)-1)$. 

\subsection{The case of long-range attraction}

Suppose first that $(g(1-h)-1)>0$. In this case there is a long range attraction force between eigenvalues. Let us rescale the density as
\begin{equation}
 \rho(x)= \mu f(x),
\end{equation} 
so that the integral equation (\ref{integral-eq}) reads
\begin{equation}
 \int \rd y f(y) K(x-y)=1-\frac{\epsilon x^2}{2}, \label{int-eq-f}
\end{equation}
where 
\be
\epsilon=\frac{\alpha}{2\pi\mu}
\ee
and it vanishes at large $\mu$. Then, from (\ref{grand-pot}) we have
\begin{equation}
 N(\mu)\equiv \frac{\partial J(\mu)}{\partial \mu}=\mu \int f(x)\rd x\equiv C\mu.
\end{equation} 
Therefore, in this theory, $\mu$ is proportional to $N$. This is again in contrast to the $\CN=3$ theories considered in \cite{dmp,mp,hklebanov}, as well as to the $\CN=2$ theories with no long-range forces analyzed above and considered previously in \cite{jkps}. Notice that the parameter $\epsilon$ can be written, at large $N$, as 
\be
\epsilon=-{\ri C \over 2 \pi} {1\over \lambda},
\ee
where
\be
\lambda={N\over k}
\ee
is the 't Hooft parameter of the theory. The M-theory limit that we are considering here ($N$ large, $k$ fixed) corresponds as usual to the strongly coupled region of the 't Hooft parameter. 
In particular, the limit of infinitely strong coupling $\lambda \rightarrow \infty$ corresponds simply to $\epsilon=0$. 
The canonical free energy in the large $N$ limit is given by
\begin{equation}
 \frac{\partial F}{\partial N}=-\mu(N),
\end{equation} 
where the function $\mu(N)$ is the inverse to $N(\mu)$, therefore
\begin{equation}
 F\approx -\frac{1}{2C} N^2.
\end{equation}

Qualitatively, the behavior of the interacting Fermi gas underlying this model is very different from the $\CN=2$ Fermi gases with no long-range forces, as the flavored theory considered above. In the model with long-range attraction, the fermions, under the action of the attractive potential, form some sort of bound state whose size is of order $\CO(1)$ at large $N$. The system is still very dense, since the density grows linearly with $N$, and therefore we expect 
the Thomas--Fermi approximation to give the right large $N$ behavior. This is very similar to the analysis of baryons at large $N$ in \cite{wittenbaryon}, whose size is of order $\CO(1)$ at large $N$ 
but whose mass grows as $\CO(N)$. In our case, the Thomas--Fermi equation is simply the standard integral equation following from the large $N$ density 
of eigenvalues of the matrix model (\ref{gad-mm}). Let us note that the bound state is formed due to the long-range attraction, even if there is no external potential and $\epsilon=0$. In this limit the integral equation has translation invariance which can be removed by assuming that the solution is centered at $0$.

The maximization principle of \cite{jafferis} says that, in order to compute $h$, we have to maximize $-F$, i.e. we have to minimize $C$. This determines the R-charge $h$ 
of the adjoint multiplets. In the 't Hooft expansion, this R-charge, in the planar limit, is a non-trivial function of the 't Hooft parameter $\lambda$. In our formalism, it is very easy to determine $h$ in the planar, strongly coupled limit $\lambda \rightarrow \infty$ (or $\epsilon=0$), since the coefficient $C|_{\epsilon=0}$ can be calculated numerically with high precision for any given $g$ and $h$. 
To do this one first solves numerically the following integral equation
\begin{equation}
\label{int-sim}
 \int f_{g,h}(y)K(x-y)\rd y=1,
\end{equation}
obtained from (\ref{int-eq-f}) by setting $\epsilon=0$. Then
\begin{equation}
 C|_{\epsilon=0}=\int f_{g,h}(x)\rd x.
\end{equation}

The large $N$, strongly coupled limit of $h$ as a function of $g$ was determined numerically in \cite{minwalla} by solving the discrete saddle-point equations for the matrix 
integral (\ref{gad-mm}) at finite $N$, $k$, and then extrapolating the result to large $N$ and $\lambda$. 
The advantage of the integral equation (\ref{int-sim}) is that it gives directly the right R-charge $h$ in the limiting region $N, \lambda \rightarrow \infty$ without the need to do an extrapolation. 
\begin{figure}
\begin{center}
\includegraphics{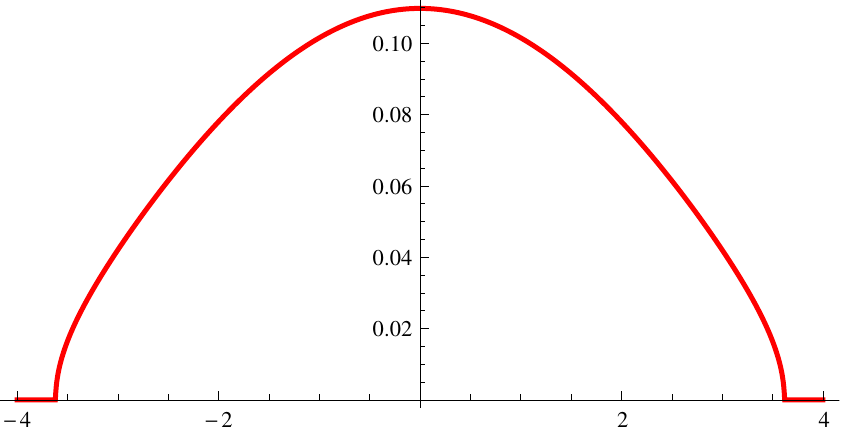}
\end{center}
\caption{The graph of $f_{g,h}(x)$, the solution of (\ref{int-sim}), for $g=2$ and $h=0.2726$.} 
\label{fgh-plot}
\end{figure}

\begin{figure}
\begin{center}
\includegraphics{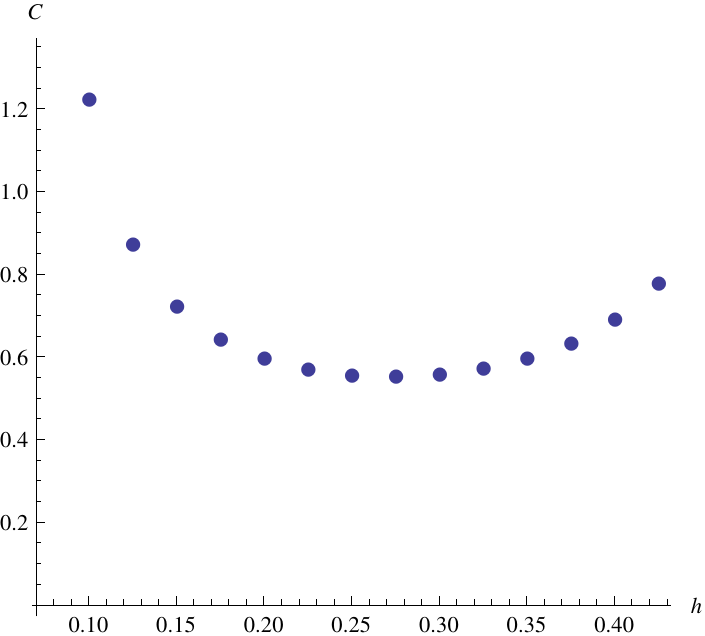}\;\;\includegraphics{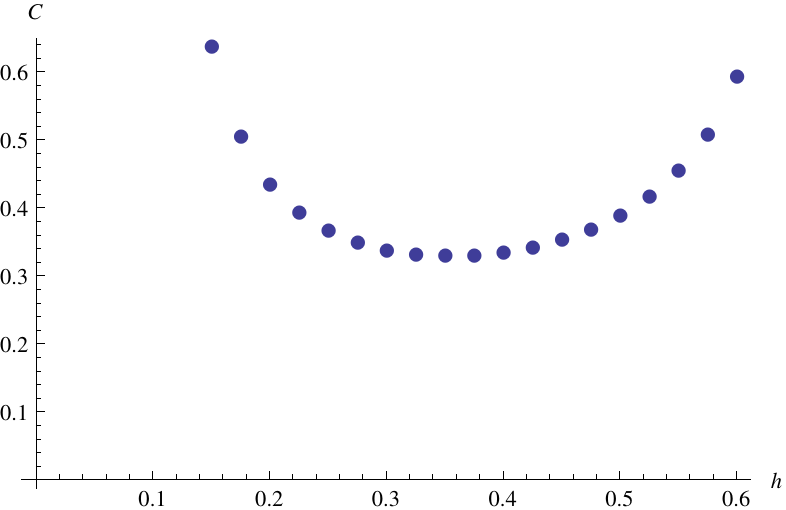}
\end{center}
\caption{The graph of $C|_{\epsilon=0}$ as a function of $h$ for $g=2$ (left) and $g=3$ (right).} 
\label{Ch-plots}
\end{figure}

An example of a numerical solution of (\ref{int-sim}) for the profile $f_{g,h}$ is shown in Fig. \ref{fgh-plot}. The behavior of $C|_{\epsilon=0}$ as a function of $h$ for $g=2,3$ is shown in Fig. \ref{Ch-plots}. Numerical calculations show that, for any real $g>1$, there is always a minimum of $C|_{\epsilon=0}$ at a certain $h_\text{min}(g)$ in the interval $0<h<1-1/g$ defined by the condition that the long range forces are attractive. Therefore 
\begin{equation}
  \lim_{g\rightarrow 1+}h_\text{min}(g)=0
\end{equation}
which agrees with the result of \cite{niarchos1,minwalla} saying that for $g=1$ the R-charge of the adjoint field tends to zero in the strong coupling limit.
Accurate numerical calculations show that $h_\text{min}(2)=0.2726\pm0.0001$ and $h_\text{min}(3)=0.3539\pm0.0001$ (see \figref{Ch-plots}) which is in agreement with the numerical values found in \cite{minwalla}.

It was pointed out in \cite{minwalla} that analytic expressions for $h$ could be obtained in the limit $g\rightarrow\infty$. We will now show how to obtain a systematic expansion for $h$ in this regime, from our integral equation (\ref{integral-eq}). In this way we will recover and extend some of the results of \cite{minwalla}. 
Let us first make the following rescaling: 
\be
x={\xi\over \sqrt{g}},\quad y={\zeta\over \sqrt{g}}, \quad f(x)=\sum_{n\geq 0}{f_n(\xi)\over g^{n+1/2}}. 
\ee
Then (\ref{int-eq-f}) reads
\begin{equation}
 \int \rd \zeta\left( \sum_{n\geq 0}\frac{f_n(\zeta)}{g^{n}}\right)\left(
A_0-\frac{2\log|\xi-\zeta|}{g}+\frac{\log g}{g}+\left(A_2-\frac{B_2}{g}\right)\frac{(\xi-\zeta)^2}{2g}+\ldots
\right)=1-\frac{\epsilon\xi^2}{2g}.
\label{int-eq-g-exp}
\end{equation} 
Let now 
\be
f_{(n)}(x)=\sum_{i=0}^nf_i(\xi)/g^{i}
\ee
be the solution at order $n$, and
\be
C_{(n)}=\int f_{(n)}(\xi)\rd\xi.
\ee
Then from (\ref{int-eq-g-exp}) at $\xi=0$ we have
\begin{equation}
 A_0C_{(n)}=1-\int \rd \zeta f_{(n-1)}(\zeta)\left(-\frac{2\log|\zeta|}{g}+\frac{\log g}{g}+\left(A_2-\frac{B_2}{g}\right)\frac{\zeta^2}{2g}+\ldots\right)\qquad \mathrm{mod}\;g^{-n-1}
\label{norm-iter}
\end{equation}
and after differentiating (\ref{int-eq-g-exp}) w.r.t. $\xi$ we get
\begin{equation}
{\rm P} \int \frac{f_{(n)}(\zeta)\rd \zeta}{\xi-\zeta}=\frac{\epsilon\xi}{2}
+\frac{A_2C_{(n)}\xi}{2}+\frac{1}{2}\int \rd \zeta f_{(n-1)}(\zeta)\left(-\frac{B_2}{g}\,\xi+\ldots\right)
\qquad \mathrm{mod}\;g^{-n-1}.
\label{eq-iter}
\end{equation}
This equation can be solved iteratively: suppose we know the solution at order $n-1$, $f_{(n-1)}(\xi)$. Then one can determine $C_{(n)}$ from (\ref{norm-iter}), and use (\ref{eq-iter}) 
to solve for $f_{(n)}$ Let us note that for any given order $n$ the r.h.s. of (\ref{eq-iter}) is a polynomial in $\xi$, and thus (\ref{eq-iter}) can be considered as the standard matrix model equation for $f_{(n)}$ with a polynomial potential which can be solved explicitly and unambiguously once $C_{(n)}$ is given.

To see how this works, let us compute the first few orders. At order $0$ we have the two equations, 
\begin{equation}
\ba
 A_0C_{(0)}&=1,\\
{\rm P} \int \frac{f_{(0)}(\zeta)\rd \zeta}{\xi-\zeta}&=\frac{\epsilon+A_2/A_0}{2}\,\xi
\ea
\end{equation}
and the solution is
\begin{equation}
 f_{(0)}(\xi)=\frac{\epsilon+A_2/A_0}{2\pi}\sqrt{\frac{4}{A_2+\epsilon A_0}-\xi^2}.
\end{equation} 
At order 1 we have
\begin{multline}
 A_0C_{(1)}=1-\frac{1}{g}\int \rd \zeta f_{(0)}(\zeta)\left(-2\log|\zeta|+\log g+A_2\,\frac{\zeta^2}{2}\right)=\\
=1-\frac{1}{A_0g}\left[\log g +1+\log\left(A_2+A_0\epsilon\right)+\frac{1}{2}\frac{A_2}{A_2+A_0\epsilon}\right]
\end{multline}
which can be rewritten as
\begin{equation}
 \frac{1}{C}=A_0g+\log g +1+\log\left(A_2+A_0\epsilon\right)+\frac{1}{2}\frac{A_2}{A_2+A_0\epsilon}+\CO\left(\frac{1}{g}\right).
\end{equation} 
Using the procedure described above up to order 2, and specializing to $\epsilon=0$, one finds
\begin{equation}
\left.\frac{1}{C}\right|_{\epsilon=0}=V_h(0)g+\log g+\frac{3}{2}+\log V_h''(0)
-\frac{1}{6gV_h''(0)}\left(1-\frac{5 V_h''''(0)}{2 V_h''(0)}\right)+\CO\left(\frac{1}{g^2}\right).
\end{equation} 
Minimizing ${C}|_{\epsilon=0}$ w.r.t. $h$ gives
\begin{equation}
 h_\text{min}=\frac{1}{2}-\frac{4}{\pi^2g}-\frac{32 \left(\pi ^2-9\right)}{3 \pi ^4 g^2}+\CO\left(\frac{1}{g^3}\right).
\end{equation} 
The first correction to the asymptotic value $1/2$ was already found in \cite{minwalla} with a related technique. 
 
\subsection{The case of long-range repulsion}
When 
\be
-\gamma\equiv(g(1-h)-1)<0
\ee
there is a long range repulsion between the fermions (or the eigenvalues of the matrix model), 
and the external potential cannot be neglected at large $N$. After rescaling (as we will see later, $\mu<0$ in this regime) 
\be
x=\sqrt{-\mu}\xi \quad y=\sqrt{-\mu}\eta, \quad \rho(x)=f(\xi)
\ee
 the equation (\ref{integral-eq}) becomes
\begin{equation}
 \gamma\int  f(\eta)|\xi-\eta|\rd\eta +\CO\left(\frac{1}{\sqrt{-\mu}}\right)=1+\frac{\alpha\xi^2}{4\pi}.
\label{repulsion-resc-eq}
\end{equation}
Suppose $f_0(\xi)$ is the solution of this equation at large $\mu$. After differentiating two times w.r.t. $\xi$, the equation (\ref{repulsion-resc-eq}) becomes 
\begin{equation}
 2\gamma f_0(\xi)=\frac{\alpha}{2\pi}.
\end{equation} 
Therefore at leading order the distribution is constant in some interval $(-\xi_0,\xi_0)$, and zero outside. The end points can be determined by evaluating (\ref{repulsion-resc-eq}) at $\xi=0$:
\begin{equation}
 \gamma\int\limits_{-\xi_0}^{\xi_0} \frac{\alpha}{4\pi\gamma}|\eta|\rd\eta=1\;\;\Longrightarrow\;\; \xi_0=\sqrt\frac{4\pi}{\alpha}.
\end{equation}
Then,
\begin{equation}
 N(\mu)\approx \sqrt{-\mu} \int f_0(\xi)\rd \xi=\sqrt{-\mu}\cdot \sqrt\frac{\alpha}{\pi\gamma^2}
\end{equation} 
and
\begin{equation}
 F\approx \frac{\pi\gamma^2 N^3}{3\alpha}. \label{repulsion-large-N}
\end{equation}
Qualitatively, in the case of repulsion, the eigenvalues spread out at large $N$ over an interval whose length grows linearly with $N$, and with constant density. 
Such configuration can be easily understood in terms of the forces between eigenvalues. Suppose we have some symmetric distribution, and let us consider an eigenvalue sitting at $x$. Let $n(x)$ be the number of eigenvalues in the interval $(-x,x)$. The long-range force between every two eigenvalues is constant and equal to $\gamma$. The total force acting on the eigenvalue at $x$ from other eigenvalues equals $\gamma n(x)$ and is directed outside of the interval $(-x,x)$. This force should be compensated by the force from the external potential, which equals %
\be
\frac{\alpha x}{2\pi}
\ee
and is directed towards the interior the interval $(-x,x)$. Then 
\be
\gamma n(x)=\frac{\alpha x}{2\pi}
\ee
and 
\be
\rho(x)=\frac{1}{2}\frac{\rd n(x)}{\rd x}=\frac{\alpha}{4\pi\gamma}.
\ee
The derivative of the free energy can be easily calculated as the energy of a probe eigenvalue added at the boundary of the distribution, and one finds
\be
\frac{\partial F}{\partial N}\approx\frac{\pi\gamma^2}{\alpha}N^2.
\ee

The result (\ref{repulsion-large-N}) can be tested in the case $g=0$. In this case, the matrix model (\ref{gad-mm}) reduces to the CS matrix model of \cite{mmcs}, and its free energy equals the free energy of CS theory on $\IS^3$ with framing $1$. The corresponding planar free energy reads (see for example eq. (4.31) in \cite{mmhouches}, to which one has to add $t^3/12$ due to framing), 
\begin{equation}
 F_0(t)=\frac{t^3}{6}-\frac{\pi^2t}{6}+\zeta(3)-\mathrm{Li}_3(\re^{-t}).
\end{equation} 
Here, $t=g_sN$ is the 't Hooft parameter and 
\be
g_s=\frac{2\pi\ri}{k}.
\ee
Then, at large $N$ and fixed $k$, 
\begin{equation}
 F(N)\approx g_s^{-2}\cdot\frac{t^3}{6}=\frac{N^3\pi\ri}{3k}
\end{equation}
which coincides with (\ref{repulsion-large-N}) in the case $g=0$ ($\gamma=1$).

\sectiono{Conclusions and open problems}

In this paper we have extended some of the results of \cite{mp} to the matrix models of $\CN=2$ CSM theories, which we have formulated in terms of interacting fermions. The resulting system can be analyzed, at large $N$, in the Hartree/Thomas--Fermi approximation, and this leads to the formulation in terms of density functionals put forward in \cite{hklebanov,jkps}. Going beyond the large $N$ approximation is in general difficult, although in the case of flavored theories with one node one can compute the next-to-leading correction by using the exchange correction to the Thomas--Fermi approximation. In the case of theories with long-range interactions, the Thomas--Fermi approximation gives an efficient way of calculating numerically the R-charge of adjoint multiplets.  

It is clear that this work leaves many open problems. One should find a more rigorous argument showing that the Thomas--Fermi approximation gives the leading contribution to the grand potential, and one should achieve a more detailed understanding of what kind of corrections are expected for $J(\mu)$, as it was done in \cite{mp} for the $\CN=3$ theories. This might lead to a way of determining, at least in the flavored theories with one node, subleading terms in $\mu$ in the grand potential, which would lead to a calculation of $1/N$ corrections to the anomalous dimensions. In the theory with long range forces, the Thomas--Fermi equation is nothing but the standard equation for the eigenvalue density, and one might try to go back to the traditional matrix model technology in order to determine $1/N$ corrections to the large $N$ result. 

\section*{Acknowledgements}
We would like to thank Thierry Giamarchi and Daniel Jafferis for useful 
conversations and correspondence. M.M. would like to thank Rico Rueedi for making his doctoral dissertation on Thomas--Fermi theory 
available to him. This work is supported by the Fonds National Suisse, subsidies 200020-126817 and 
200020-137523. P.P. is also supported by FASI RF 14.740.11.0347.


\begin{thebibliography}{99}
\bibliographystyle{plain}

 \bibitem{abjm}
 O.~Aharony, O.~Bergman, D.~L.~Jafferis and J.~Maldacena, ``N=6 superconformal Chern-Simons-matter theories, M2-branes and their gravity duals,''
  JHEP {\bf 0810}, 091 (2008)
  [arXiv:0806.1218 [hep-th]].
  
\bibitem{amariti}
 A.~Amariti, ``On the exact R charge for N=2 CS theories,''
  JHEP {\bf 1106}, 110 (2011)
  [arXiv:1103.1618 [hep-th]].
  
\bibitem{benini}
F.~Benini, C.~Closset and S.~Cremonesi, ``Chiral flavors and M2-branes at toric CY4 singularities,''
  JHEP {\bf 1002}, 036 (2010)
  [arXiv:0911.4127 [hep-th]].

 
\bibitem{k1} 
S.~Cheon, H.~Kim, N.~Kim, ``Calculating the partition function of N=2 Gauge theories on $S^3$ and AdS/CFT correspondence,''
  JHEP {\bf 1105}, 134 (2011).
  [arXiv:1102.5565 [hep-th]]. 

\bibitem{cmp}
R.~Couso Santamar\'\i a, M.~Mari\~no and P.~Putrov, ``Unquenched flavor and tropical geometry in strongly coupled Chern-Simons-matter theories,''
  JHEP {\bf 1110}, 139 (2011)
  [arXiv:1011.6281 [hep-th]].
  
 \bibitem{dmp}
 N.~Drukker, M.~Mari\~no, P.~Putrov, ``From weak to strong coupling in ABJM theory,''
  Commun.\ Math.\ Phys.\  {\bf 306}, 511-563 (2011).
  [arXiv:1007.3837 [hep-th]]. 

\bibitem{fhm}
H.~Fuji, S.~Hirano, S.~Moriyama, ``Summing Up All Genus Free Energy of ABJM Matrix Model,''
  JHEP {\bf 1108}, 001 (2011).
  [arXiv:1106.4631 [hep-th]].

\bibitem{gy}
 D.~Gaiotto and X.~Yin, ``Notes on superconformal Chern-Simons-Matter theories,''
  JHEP {\bf 0708}, 056 (2007)
  [arXiv:0704.3740 [hep-th]].
  

\bibitem{hama}
N.~Hama, K.~Hosomichi, S.~Lee, ``Notes on SUSY Gauge Theories on Three-Sphere,''
  JHEP {\bf 1103}, 127 (2011).
  [arXiv:1012.3512 [hep-th]].

  
 \bibitem{hklebanov}
 C.~P.~Herzog, I.~R.~Klebanov, S.~S.~Pufu, T.~Tesileanu, 
 ``Multi-Matrix Models and Tri-Sasaki Einstein Spaces,''
  Phys.\ Rev.\  {\bf D83}, 046001 (2011).
  [arXiv:1011.5487 [hep-th]].


  \bibitem{hillery}
 M.~Hillery, R.~F.~O'Connell, M.~O.~Scully, E.~P.~Wigner, ``Distribution functions in physics: Fundamentals,''
  Phys.\ Rept.\  {\bf 106}, 121-167 (1984).

\bibitem{quiver1}
  Y.~Imamura, K.~Kimura, ``On the moduli space of elliptic Maxwell-Chern-Simons theories,''
  Prog.\ Theor.\ Phys.\  {\bf 120}, 509-523 (2008).
  [arXiv:0806.3727 [hep-th]].



\bibitem{jaff-flav}
D.~L.~Jafferis, ``Quantum corrections to N=2 Chern-Simons theories with flavor and their AdS(4) duals,''
  arXiv:0911.4324 [hep-th].

\bibitem{jafferis}
D.~L.~Jafferis, ``The Exact Superconformal R-Symmetry Extremizes Z,''
  JHEP {\bf 1205}, 159 (2012)
  [arXiv:1012.3210 [hep-th]].

\bibitem{jkps}
 D.~L.~Jafferis, I.~R.~Klebanov, S.~S.~Pufu, B.~R.~Safdi, ``Towards the F-Theorem: N=2 Field Theories on the Three-Sphere,''
  JHEP {\bf 1106}, 102 (2011).
  [arXiv:1103.1181 [hep-th]].


\bibitem{ghp2} 
  D.~R.~Gulotta, C.~P.~Herzog and S.~S.~Pufu,
  ``Operator Counting and Eigenvalue Distributions for 3D Supersymmetric Gauge Theories,''
  JHEP {\bf 1111}, 149 (2011)
  [arXiv:1106.5484 [hep-th]].

\bibitem{ghp1} 
  D.~R.~Gulotta, C.~P.~Herzog and S.~S.~Pufu,
  ``From Necklace Quivers to the F-theorem, Operator Counting, and T(U(N)),''
  JHEP {\bf 1112}, 077 (2011)
  [arXiv:1105.2817 [hep-th]].
  
\bibitem{quiver2}
  D.~L.~Jafferis, A.~Tomasiello, ``A Simple class of N=3 gauge/gravity duals,''
  JHEP {\bf 0810}, 101 (2008).
  [arXiv:0808.0864 [hep-th]].


  \bibitem{kwy}
A.~Kapustin, B.~Willett and I.~Yaakov, ``Exact Results for Wilson Loops in Superconformal Chern-Simons Theories with Matter,''
  JHEP {\bf 1003}, 089 (2010)
  [arXiv:0909.4559 [hep-th]].

\bibitem{kmss}
A. Klemm, M. Mari\~no, M. Schiereck and M. Soroush, ``ABJM Wilson loops in the Fermi gas approach," to appear. 

\bibitem{kunz}
H. Kunz and R. Rueedi, ``Atoms and quantum dots with a large number of electrons: the ground-state energy," Phys. Rev. A {\bf 81}, 032122 (2010). 

\bibitem{lr}
R. Lawrence and L. Rozansky, ``Witten-Reshetikhin-Turaev invariants
of Seifert manifolds,'' Comm. Math. Phys.
{\bf 205} (1999) 287.

\bibitem{lieb}
 E.~H.~Lieb, ``Thomas--Fermi and related theories of atoms and molecules,''
  Rev.\ Mod.\ Phys.\  {\bf 53}, 603 (1981)
  [Erratum-ibid.\  {\bf 54}, 311 (1982)].
 
 \bibitem{mmcs}
  M.~Mari\~no, ``Chern-Simons theory, matrix integrals, and perturbative three manifold invariants,''
  Commun.\ Math.\ Phys.\  {\bf 253}, 25 (2004)
  [hep-th/0207096].
  
  \bibitem{mmhouches}
  M.~Mari\~no, ``Les Houches lectures on matrix models and topological strings,''
  hep-th/0410165.
  
 \bibitem{lectures}
 M.~Mari\~no, ``Lectures on localization and matrix models in supersymmetric Chern-Simons-matter theories,'' J.\ Phys.\ A A {\bf 44}, 463001 (2011)
  [arXiv:1104.0783 [hep-th]].

\bibitem{mpwilson}
M.~Mari\~no and P.~Putrov, ``Exact Results in ABJM Theory from Topological Strings,''
  JHEP {\bf 1006}, 011 (2010)
  [arXiv:0912.3074 [hep-th]].
  
  \bibitem{mp}
   M.~Mari\~no, P.~Putrov, ``ABJM theory as a Fermi gas,'' J.\ Stat.\ Mech.\  {\bf 1203}, P03001 (2012)
  [arXiv:1110.4066 [hep-th]].
 
 \bibitem{ms}
 D.~Martelli, J.~Sparks, ``The large N limit of quiver matrix models and Sasaki-Einstein manifolds,''
  Phys.\ Rev.\  {\bf D84}, 046008 (2011).
  [arXiv:1102.5289 [hep-th]]. 
  
 \bibitem{minwalla}
 S.~Minwalla, P.~Narayan, T.~Sharma, V.~Umesh and X.~Yin, ``Supersymmetric States in Large N Chern-Simons-Matter Theories,''
  JHEP {\bf 1202}, 022 (2012)
  [arXiv:1104.0680 [hep-th]].

\bibitem{mn}
 T.~Morita and V.~Niarchos, ``F-theorem, duality and SUSY breaking in one-adjoint Chern-Simons-Matter theories,''
  Nucl.\ Phys.\ B {\bf 858}, 84 (2012)
  [arXiv:1108.4963 [hep-th]].
  
\bibitem{no}
J. Negele and H. Orland, {\it Quantum many-particle systems}, Westview Press, 1998. 

\bibitem{niarchos}
 V.~Niarchos, ``Comments on F-maximization and R-symmetry in 3D SCFTs,''
  J.\ Phys.\ A A {\bf 44}, 305404 (2011)
  [arXiv:1103.5909 [hep-th]].

\bibitem{niarchos1} 
  V.~Niarchos,
  ``R-charges, Chiral Rings and RG Flows in Supersymmetric Chern-Simons-Matter Theories,''
  JHEP {\bf 0905}, 054 (2009)
  [arXiv:0903.0435 [hep-th]].

\bibitem{pestun}
 V.~Pestun, ``Localization of gauge theory on a four-sphere and supersymmetric Wilson loops,''
  Commun.\ Math.\ Phys.\  {\bf 313}, 71 (2012)
  [arXiv:0712.2824 [hep-th]].
 
 \bibitem{suyama}
  T.~Suyama, ``On Large N Solution of Gaiotto-Tomasiello Theory,''
  JHEP {\bf 1010}, 101 (2010)
  [arXiv:1008.3950 [hep-th]].
  
\bibitem{tomi}
Y. Tomishima, ``The Thomas-Fermi Theory at Finite Temperature Including WeizsŠcker and Correlation Corrections: Temperature Green's Function Formalism," J. Phys. Soc. Jpn. {\bf 54} (1985) 1282.

\bibitem{wittenbaryon}
  E.~Witten, ``Baryons in the $1/N$ Expansion,''
  Nucl.\ Phys.\ B {\bf 160}, 57 (1979).
  
\bibitem{cz}
C. K. Zachos, D. B. Fairlie and T. L. Curtright (eds.), {\it Quantum Mechanics in phase space}, World Scientific, Singapore, 2005.
  
\end{thebibliography}
\end{document}